\documentclass[aps,physrev,reprint,amsmath,amssymb]{revtex4-2}
\pdfoutput=1

\usepackage{graphicx}
\graphicspath{{./img/}{./}}

\usepackage{hyperref}

\newcommand{\td}{\mathrm{d}}
\newcommand{\ti}{\mathrm{i}}

\newcommand{\Tr}{\mathrm{Tr}}

\begin{document}

\title{Charged and neutral fixed points in the $O(N) \oplus O(N)$ model with Abelian gauge fields}

\author{Aron J. Beekman}
\email{aron@phys-h.keio.ac.jp}
\affiliation{Department of Physics, and Research and Education Center for Natural Sciences, Keio University, 3-14-1 Hiyoshi, Kohoku-ku, Yokohama 223-8522, Japan}
\author{Gergely Fej\H{o}s}
\email{fejos@keio.jp}
\affiliation{Department of Physics, Keio University, 3-14-1 Hiyoshi, Kohoku-ku, Yokohama 223-8522, Japan}
\affiliation{Institute of Physics, E\"otv\"os University, 1117 Budapest, Hungary}

\date {\today}
\begin{abstract}
 In the Abelian-Higgs model, or Ginzburg-Landau model of superconductivity, the existence of an infrared stable charged fixed point ensures that there is a parameter range where the superconducting phase transition is second order, as opposed to fluctuation-induced first order as one would infer from the Coleman-Weinberg mechanism. We study the charged and neutral fixed points of a two-field generalization of the Abelian-Higgs model, where two $N$-component fields are coupled to two gauge fields and to each other, using the functional renormalization group. Focusing mostly on three dimensions, in the neutral case, this is a model for two-component Bose-Einstein condensation, and we confirm the fixed-point structure established in earlier works using different methods. The charged model is a dual theory of two-dimensional dislocation-mediated quantum melting. We find the existence of three charged fixed points for all $N>2$, while there are additional fixed points for $N=2$.
\end{abstract}

\maketitle

\section{Introduction}

The study of the nature of the superconducting phase transition has an interesting history. As physicists became aware of renormalization and universality, they found that the behavior of the critical point related to the second-order phase transition of a neutral superfluid was significantly changed compared to a simple mean-field approximation, culminating in the sharp predictions of what is now called the Wilson-Fisher fixed point of $O(N)$ models. Conversely, while charged superconductors not only have a dynamical order parameter field but also are coupled to dynamical gauge fields, real-world superconductors follow the predictions from mean-field theory remarkably well. It was realized by Ginzburg that the temperature range where fluctuations are relevant, is actually very narrow. Meanwhile it was established that the dynamical gauge field may preclude a second-order phase transition entirely, instead leading to a fluctuation-induced first-order transition (via the Coleman-Weinberg mechanism). This is now textbook material; see, for instance, Ref.~\cite{Herbut07}.

\begin{figure}[b]
\centering
\includegraphics[width=.35\textwidth]{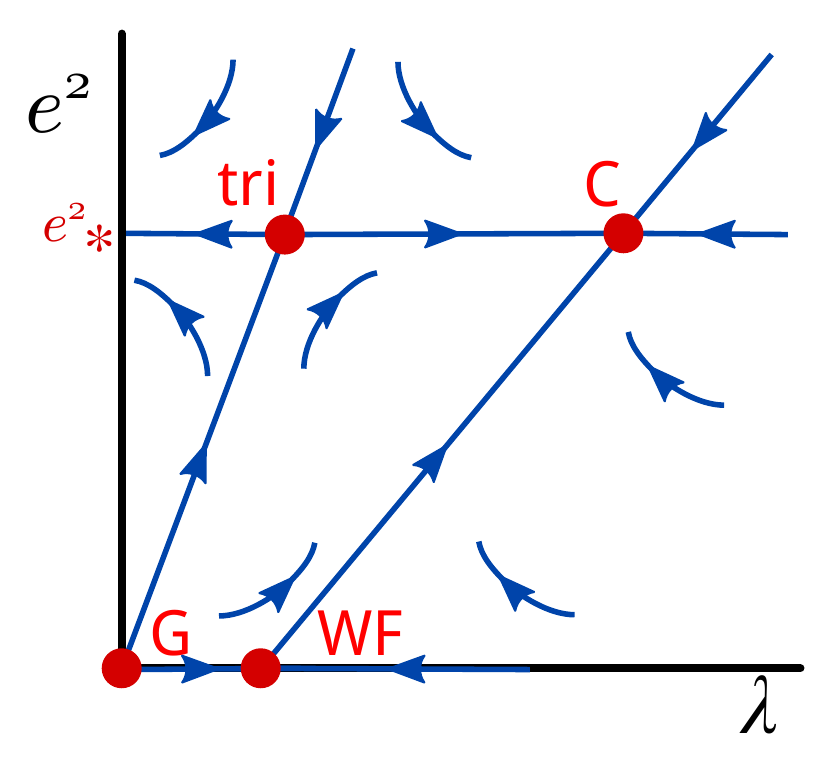}
 \caption{Schematic flow diagram of the one-field Abelian-Higgs model in $d=3$. Next to the usual Gaussian (G) and Wilson-Fisher (WF) neutral fixed points, there are two charged fixed points: one infrared stable (C) and the other tricritical (tri), which separates between regions at high $\kappa^2 = \lambda / e^2$ with a second-order phase transition, and at low $\kappa^2$ with a first-order phase transition.}
 \label{fig:AH flow diagram}
 \end{figure}

Later Kleinert put forward, on the basis of a superfluid-superconductor duality mapping, that the nature of the phase transition depends on the ratio $\kappa = \lambda_\mathrm{L} / \xi$, where $\lambda_\mathrm{L}$ is the London penetration depth and $\xi$ is the coherence length~\cite{Kleinert83}.
In terms of the renormalization group parameters, $\kappa^2 \propto \lambda/e^2$, where $\lambda$ is the order parameter self-coupling and $e$ is the electric charge [see Eq.~\eqref{eq:AH action}].
There is a critical value $\kappa_\mathrm{c} \approx 0.8/\sqrt{2}$ that separates between superconductors with either first-order (low $\kappa$) or second-order (high $\kappa$) phase transitions. Note that this almost coincides with the distinction between type-I and type-II superconductors at $\kappa = 1/\sqrt{2}$ referring to their properties under an applied magnetic field. In renormalization group terms, the second-order phase transition takes place at an infrared stable {\em charged fixed point} where $e^2_* \neq 0$. 

While numerical calculations verified this state of affairs, renormalization group (RG) derivations based on the $\epsilon$ expansion or the $1/N$ expansion could confirm the existence of the charged fixed point only at very high $N$~\cite{Herbut07}. However, a field-theoretical RG calculation directly in $d=3$~\cite{Herbut96}  and functional renormalization group (FRG) calculations~\cite{Bergerhoff96} could reproduce the flow diagram down to $N=2$, although these works had their own limitations. Most recently, in Refs.~\cite{FejosHatsuda16,FejosHatsuda17} it was shown that the FRG can fully reproduce the flow diagram depicted in Fig.~\ref{fig:AH flow diagram} for all $N$. There are four fixed points in total. The two neutral fixed points at $e^2_* = 0$ correspond to the usual Gaussian and Wilson-Fisher fixed points of the neutral $O(N)$ model. They are unstable against any finite value of the charge. There are two charged fixed points, at the same value of the charge $e^2_* > 0$. There is an infrared stable fixed point at higher $\lambda$ which corresponds to the second-order superconducting phase transition. The fixed point at lower $\lambda$ is tricritical and separates between regions where the phase transition is first and second order.  This shows that in models that contain dynamical gauge fields, the existence of charged fixed points indicates the possibility of having second-order phase transitions instead of fluctuation-induced first-order ones.

Here we study the fixed point structure of a generalization of the Abelian-Higgs model (or Ginzburg-Landau model) to two $N$-component fields which are each coupled to a dynamic $U(1)$-gauge field and to each other. The motivation for undertaking this work is threefold. First, it is interesting to expand the method of Refs.~\cite{FejosHatsuda16,FejosHatsuda17} to other models in the search for yet undiscovered charged fixed points.

Second, since the early days there has been interest in the critical properties of the (uncharged) $O(N_1) \otimes O(N_2)$ and $O(N_1) \oplus O(N_2)$ models, mainly due to the presence of multicritical points, where more than one coupling constant is turned to a critical value and more than two phases coexist (like the tricritical point in Fig.~\ref{fig:AH flow diagram}). Early work focused on multicriticality in anisotropic antiferromagnets~\cite{Kosterlitz76}. This has found a recent application for $N_1=N_2\equiv N=2$ in describing two-component Bose-Einstein condensates (BECs)~\cite{Ceccarelli15,Ceccarelli16}. Here two order parameters are coupled to each other. When condensation takes place, if the coupling is weak, both components condense at the same time as if they were not coupled at all; this is called the miscible phase. Conversely, if repulsive coupling is strong, only one field condenses; this is called the immiscible phase. In renormalization group terms, in the first case the system flows to a decoupled fixed point where the coupling vanishes, while in the second case it flows to a finite positive value of the coupling. Here we wish to reproduce the results of the neutral model in a simpler way, and we study how the fixed point structure is modified when coupled to dynamical gauge fields.

Third, the charged $O(2) \oplus O(2)$ model is a toy model to study the solid-to-hexatic melting quantum phase transition in two spatial dimensions, using the standard $d$-dimensional quantum-to-$d+1$-dimensional classical mapping. Using a generalization of Abelian-Higgs or vortex-boson duality, the two phonons of a $d=2$ solid are mapped to two free and massless vector gauge fields, which mediate interactions between dislocation topological defects~\cite{Zaanen04,Beekman17}. In a quantum version of Kosterlitz-Thouless-Halperin-Nelson-Young dislocation-mediated melting, the dislocations can Bose condense to restore translational symmetry. Just like the two-component Bose-Einstein condensates, we have situations where dislocations restore symmetry in only one or in both directions---corresponding to the immiscible and miscible phases. These phases then have the symmetry of smectic and hexatic liquid crystals, respectively. Recent experiments have shown evidence of a possible quantum hexatic liquid crystal phase in helium monolayers on a graphite substrate~\cite{Nakamura16}. It is our ultimate goal to extract critical exponents related to the quantum critical point of this solid-to-hexatic phase transition, such that experiments can confirm whether the quantum hexatic phase exists. Therefore, we need to understand how the critical behavior is affected by the addition of gauge fields (representing the long-range interaction between bosons---the dislocations). As a first step we here investigate a simplified model, which differs from the full theory of Ref.~\cite{Beekman17} in two respects: First, longitudinal and transverse phonons have different dynamics, but here we look at two identical gauge fields only. Second,  dislocation dynamics obeys the so-called glide constraint, which effectively decouples the longitudinal phonon from the dislocation condensation. We leave these two complications for future consideration. In light of this motivation, our main interest is in $N=2$ and $d=3$, although we study other cases as well.

We use an implementation of the FRG, building upon the method developed in Refs.~\cite{FejosHatsuda16,FejosHatsuda17}. What distinguishes the FRG from the ordinary perturbative Wilsonian renormalization group is that, not only are the couplings flowing under change of the momentum scale $k$, but so is the effective action itself, and furthermore, one is also free to choose between regularization schemes. The advantage of the FRG compared to the field-theoretical renormalization group is that results can be extracted for all dimensions $2<d<4$. While we provide some explanation of our method in Sec.~\ref{sec:Corrections to the O(N)-model}, we refer to the reviews in Refs.~\cite{Berges02,Kopietz10} for details about the FRG in general, and to Refs.~\cite{FejosHatsuda16,FejosHatsuda17} for details of the present method including a consistent gauge-fixing scheme. One modification that we implement is that we include a correction from terms of one order higher in the fields, based on Ref.~\cite{PapenbrockWetterich95}. This is explained in Sec.~\ref{sec:Corrections to the O(N)-model}. 

\begin{table*}
 \begin{tabular}{lcccccc}
  this work & $m^2$ & $\lambda$ & $u$ & $g$ & $w$ &$e$ \\
  \hline
  Ref.~\cite{PapenbrockWetterich95}, Eq.(3.5) & $\kappa$ & $\lambda$ & $u_3$ & & &\\
  Ref.~\cite{Ceccarelli15}, Eq.(3) & $r$ & $12g$ &  & $u$ & &\\
  Ref.~\cite{Ceccarelli16}, Eq.(16) & $r$ & $\tfrac{1}{2}v$ & & $\tfrac{1}{8}u$ & &\\
  Ref.~\cite{Eichhorn13}, Eq.(1) & $r_\phi = r_\chi$ & $u_\phi = u_\chi$ & & $\frac{1}{12}u_{\phi\chi}$ & &\\
  Ref.~\cite{Abreu08}, Eq.(1) & $r_{01} = r_{02}$ & $2 u_{01} = 2u_{02}$ & & $\tfrac{1}{6}u_{03}$ & & $e_{01} = e_{02}$\\
  Ref.~\cite{Sakhi13}, Eq.(1) & $r_1=r_2$ & $2 u = 2v$ & & $\tfrac{1}{6}w$ & & $e_a=e_b$ \\
  Ref.~\cite{Sakhi18}, Eq.(1) &$r_1=r_2$ & $\tfrac{6}{N}\lambda_1 = \tfrac{6}{N}\lambda_2$ & $\tfrac{4}{N^2}g_1 =  \tfrac{4}{N^2}g_2$ & $\tfrac{1}{2N}\lambda_3$ & $\tfrac{6}{N^2}g_3 = \tfrac{6}{N^2}g_4$ & $1$
 \end{tabular}
 \caption{Correspondences of coupling constants in this work [see Eq.~\eqref{eq:2field AH Lagrangian} for the most general form] to those of related works.}\label{table:coupling constants}
\end{table*}

Now we discuss the relation to other works. Neutral $O(N_1) \oplus O(N_2)$ models have been studied since the 1970s~\cite{Kosterlitz76}. A large body of work including Refs.~\cite{Pelissetto02,Pelissetto05,Pismak09,Calabrese03,Calabrese04} using various RG schemes ($\epsilon$ expansion, $1/N$ expansion, minimal subtraction, field-theoretical RG) has established the existence of six fixed points near $d=4$, which reduce to four if $N_1 = N_2$ and furthermore if the coupling constants of the two fields are identical. In $d=3$ there are four fixed points that explain the physics of two-component Bose-Einstein condensates~\cite{Ceccarelli15,Ceccarelli16}. Some early work using the FRG was carried out in Refs.~\cite{Bornholdt95,Bornholdt96}. The work closest to our method is that of Refs.~\cite{Eichhorn13,Borchardt16}, also using the FRG.  We confirm these earlier results for the neutral $O(N)\oplus O(N)$ models.

We are aware of only a little work on coupling two scalar fields to gauge fields. Reference ~\cite{Abreu08} studies a two-field model coupled to a single gauge field in the large-$N$ limit. A self-dual model of Josephson junction arrays was studied, also in the large-$N$ limit, in Refs.~\cite{Sakhi13,Sakhi18}, where two fields were coupled to two gauge fields, but these gauge fields were also coupled to each other through a mutual Chern-Simons term. This should reduce to the model studied here when this mutual Chern-Simons coupling vanishes. These works seem to find charged fixed points at larger values of $N$. Here we establish the existence of charge fixed points down to the lowest $N$. Recently, Refs.~\cite{Benvenuti18,Benvenuti19} studied charged and neutral fixed points of two-field theories in the $1/N$ and $\epsilon$ expansions, in the context of boson-fermion dualities.

In order to facilitate the comparison to results obtained by other authors, we will use the following notation. For the scalar fields, $N$ denotes the number of real components, while $M$ denotes the number of complex components, such that the correspondence is $M = 2N$. We always work in $d$-dimensional Euclidean space, which describes either a finite-temperature $d$-dimensional system with a thermal phase transition, or a Wick rotation of a $d-1$-dimensional zero-temperature quantum system with a quantum phase transition. The correspondence of our notation to that of other works is detailed in Table~\ref{table:coupling constants}.

The outline of this paper is as follows. In order to understand the procedure that we are using for the two-field model, in Sec.~\ref{sec:Corrections to the O(N)-model} we, in part, apply the method of Ref.~\cite{PapenbrockWetterich95} to the neutral $O(N)$ model, and we show how this gives corrections to the flow equations for the self-coupling. This also establishes our implementation of the FRG. In Sec.~\ref{sec:Neutral two-field model} we treat the neutral two-field model using this technique and discuss the fixed-point structure that arises. In Sec.~\ref{sec:Charged two-field model} we finally extend the Abelian-Higgs model of Refs.~\cite{FejosHatsuda16,FejosHatsuda17} to two scalar fields coupled to two gauge fields, and we discuss the fixed-point structure compared both to the single-field charged model (Abelian-Higgs) and to the neutral two-field model. We discuss the relevance of these results as well as an outlook to future work in Sec. ~\ref{sec:Conclusions}.

\section{Corrections to the $O(N)$-model}\label{sec:Corrections to the O(N)-model}
The archetype for studying phase transitions is the $O(N)$ model of an $N$-component real field $\phi_a$ with mass $m^2$ and self-coupling $\lambda$. Early works established the existence of a nontrivial fixed point at finite $\lambda$, now called the Wilson-Fisher fixed point. One of the advantages of the FRG is that this result can be reproduced very quickly~\cite{Berges02,Kopietz10}. Furthermore, Papenbrock and Wetterich showed how a two-loop result can be reproduced~\cite{PapenbrockWetterich95}. Here we implement a simplified version of this method, moreover, using a Litim-type regulator~\cite{Litim00} that was not available at the time, which we shall derive here in some detail as a warm-up for the calculations to follow.

The first part of the method of Ref.~\cite{PapenbrockWetterich95} is to include a term of sixth order in $\phi_a$:
\begin{equation}\label{eq:O(N) model sixth order}
 \mathcal{S} = \int \td^d x \Big(\frac{1}{2} (\nabla \phi_a)^2 +  \frac{m^2}{2} \phi_a^2 + \frac{\lambda}{4!} (\phi_a^2)^2 + \frac{u}{48} (\phi_a^2)^3\Big).
\end{equation}
It is of course known that the coupling $u$ is irrelevant and that this term is usually neglected. However, the flow of $\lambda$ will now contain a contribution proportional to $u$, and by solving the flow equation of the (irrelevant) coupling $u$, we can substitute this solution into the flow equation for $\lambda$ to obtain a higher-order correction. Reference ~\cite{PapenbrockWetterich95} furthermore implements a sophisticated scheme for handling wave function renormalization, but for our purposes that is not needed, and we do without this refinement.

The basis of the functional or exact renormalization group (the FRG) is the Wetterich equation in momentum space~\cite{Wetterich93}
 \begin{equation}\label{eq:Wetterich equation}
  \partial_k \Gamma_k = \frac{1}{2} \Tr \int \td^d p \td^d q \; 
  \frac{ \partial_k R_k (-q,-p)}{
  \left( \Gamma^{(2)}_k + R_k \right)_{ab} (p,q)}.
 \end{equation}
 Here $\Gamma_k[\phi]$ is the regulated effective action where $k$ is the momentum scale, the fluctuations below which are suppressed. We have $\lim_{k\to0} \Gamma_k = \Gamma$, the effective action. This is done by by adding a term $\int \phi_a (R_k)_{ab} \phi_b$ to the action, where $R_k$ is called the {\em regulator}, and crudely speaking acts as a scale-dependent mass term that prevents the occurrence of any infrared divergence. Furthermore, $(\Gamma^{(2)}_k)_{ab} = \delta^2 \Gamma_k / \delta \phi_a \delta \phi_b$ is the second-derivative matrix, or the propagator for the flow of the effective action itself. The matrix trace $\Tr$ is over the $ab$ indices. For more details on the FRG, see the reviews in Refs.~\cite{Berges02,Kopietz10}.
 
 For our case, Eq.~\eqref{eq:O(N) model sixth order}, the scale-dependent effective action in the local potential approximation~\cite{Kopietz10} is
 \begin{equation}\label{eq:O(N) model sixth order effective action}
  \Gamma_k = \int \td^d x \Big( \frac{\phi_a}{2} (-\nabla^2 + m_k^2) \phi_a + \frac{\lambda_k}{4!} (\phi_a^2)^2 + \frac{u_k}{48} (\phi_a^2)^3 \Big),
 \end{equation}
where the effective potential is expanded up to sixth order in the fields. The reason for this approximation is that in principle we are looking for fixed-point solutions in $d=3$, and we wish to keep relevant and marginal interactions only. More sophisticated truncations take into account the anomalous dimension of the field and include the full field dependence of the wave function renormalization factors~\cite{Bartosch13}. This method hints at the existence of an irrelevant eigenvalue of small magnitude that can lead to a change in scaling. As a first approximation, however, here we neglect all scale dependence of the wave function renormalizations, and leave exploration of the effects of the anomalous dimensions for further studies.
 
 The flows of the now scale-dependent couplings are obtained by explicitly calculating the left- and the right-hand side of Eq.~\eqref{eq:Wetterich equation} and comparing the coefficients of each power of the fields. This process can typically be eased by going to momentum space and by evaluating with respect to constant background fields, that is, by setting all fields to a uniform, constant value after having derived the matrix $(\Gamma^{(2)}_k)_{ab}$. For our case, due to the $O(N)$ symmetry of Eq.~\eqref{eq:O(N) model sixth order effective action}, we are at liberty to set the constant field to have only the first component nonzero: $\phi_a(x) = \phi_{0a} = \delta_{a1} \phi_0$. Throughout this paper the $_0$ label indicates a constant field. We then obtain the matrix elements
 \begin{align}\label{eq:ON propagator matrix}
  \Gamma^{(2)}_{k,ab}(p,q) &= \Big( (q^2 + m_k^2)\delta_{ab} + \frac{1}{3!} \lambda_k {\phi_0}^2 ( 2 \delta_{a,1} \delta_{b,1} +  \delta_{ab}) \nonumber\\
  &\phantom{mm} + \frac{1}{8}  u_k  {\phi_0}^4( 4 \delta_{a,1} \delta_{b,1} +  \delta_{ab})\Big)\delta(p+q) .
  \end{align}
We add the regulator devised by Litim~\cite{Litim00}, which is different from the regulator used in Ref.~\cite{PapenbrockWetterich95}:
 \begin{align}
  R_{k,ab} (p,q) &= R_{k,ab}(q) \delta^d (p+q) (2\pi)^d,\label{eq:Litim regulator} \\
  R_{k,ab}(q) &= R_k(q) \delta_{ab} =  (k^2 - q^2) \Theta(k^2 - q^2) \delta_{ab}.\label{eq:Litim regulator diagonal}
 \end{align}
 Here $\Theta(x)$ is the Heaviside step function.  With this regulator the Wetterich equation becomes
 \begin{equation}\label{eq:Wetterich equation Litim}
    \partial_k \Gamma_k = \frac{1}{2} \Tr  \int \td^d q \; \left( \Gamma^{(2)}_k + R_k \right)_{ab}^{\phantom{ab}-1} (q,-q) 2k \Theta(k^2 - q^2).
 \end{equation}
In our case the matrix is diagonal. Here and below the equation is evaluated for constant field configurations; in this case the terms $\left( \Gamma^{(2)}_k + R_k \right)_{ab}^{\phantom{ab}-1}$ contain a trivial delta function $\delta(0)$ which corresponds just to the spacetime volume and is canceled against a similar term on the left-hand side. The integration over $q$ can be easily evaluated. Comparing terms on the left- and right-hand sides gives the flow equations
\begin{widetext}
\begin{align}
 k \partial_k m^2_k &= - \frac{\Omega_d}{d} k^{d+2} \frac{\lambda_k (N+2)}{3(k^2 + m^2_k)^2},\\
 k \partial_k \lambda_k &= \frac{\Omega_d}{d} k^{d+2}   \frac{2 \lambda_k^2 (N+8) - 9 u_k (k^2+m_k^2)(N+4)}{3(k^2 + m^2_k)^3},\label{eq:one-field flow lambda}\\
 k \partial_k u_k &= -\frac{\Omega_d}{d} k^{d+2}   \frac{2 \lambda_k \left( \lambda_k^2 (N+26) - 9 u_k (k^2 + m_k^2)(N+14) \right)}{3(k^2 + m^2_k)^4}.
\end{align}
Here $\Omega_d = \int \td \Omega_d/(2\pi)^d = 2/[(4\pi)^{d/2} \Gamma(d/2)]$ is the integral over the solid angle of a $d$-dimensional hypersphere. Note that $N$ comes in only from summing over $N-1$ identical contributions $a=b>1$ of Eq.~\eqref{eq:ON propagator matrix}.

We are interested in the flows of dimensionless couplings. We therefore introduce $\bar{m}^2_k = k^{-2} m^2_k$, $\bar{\lambda}_k = k^{d-4} \lambda_k$, $\bar{u}_k = k^{2d-6} u_k$. This leads to the final form of the flow equations :
\begin{align}
 k \partial_k \bar{m}^2_k &= - 2 \bar{m}^2_k -\frac{\Omega_d}{d}  \frac{\bar{\lambda}_k (N+2)}{3(1 + \bar{m}^2_k)^2},\label{eq:one-field flow m}\\
 k \partial_k \bar{\lambda}_k &=  (d-4)\bar{\lambda}_k+ \frac{\Omega_d}{d}\frac{2 \bar{\lambda}^2_k (N+8)- 9 \bar{u}_k(1+\bar{m}^2_k) (N+4)}{3(1 + \bar{m}^2_k)^3},\\
 k \partial_k \bar{u}_k &=  (2d-6)\bar{u}_k -\frac{\Omega_d}{d}\frac{2 \bar{\lambda}\left( \bar{\lambda}^2 (N+26) - 9 \bar{u}_k(1+\bar{m}^2_k)(N+14) \right)}{3(1 + \bar{m}^2_k)^4}.\label{eq:one-field flow u}
 \end{align}
The fixed points of this model are the points where these flows vanish simultaneously. As we mentioned above, the next-order correction is obtained by solving the equation for $\bar{u}$ in terms of  $\bar{m}^2_k$ and $\bar{\lambda}_k$, and substituting it into the flow equation for $\bar{\lambda}_k$:
\begin{equation}
 \bar{u}_* = \frac{\bar{\lambda}_k^3 \Omega_d (N + 26)}{9(1 + \bar{m}_k^2 ) \left( (d-3)d (1 + \bar{m}_k^2 )^3 + \bar{\lambda}_k \Omega_d(14+N) \right) }.
\end{equation}
\end{widetext}
Equation ~\eqref{eq:one-field flow m} is independent of $\bar{u}_k$ and, equating it to zero, can be solved for $\bar{m}^2_k$ as a function of $\bar{\lambda}_k$. For general $\alpha$'s the solution to
\begin{equation}
 - x - \frac{\alpha}{(1+x)^2} = 0
\end{equation}
is
\begin{equation}\label{eq:flow m general solution}
 x = - \frac{4}{3} \sin^2 \big(\frac{1}{3} \sin^{-1} \tfrac{1}{2}\sqrt{27 \alpha} \big).
\end{equation}
In our case $\alpha = \frac{\Omega_d}{6d} (N+2)\bar{\lambda}_k $, and we find that
\begin{equation}
 \bar{m}^2_{*} = -\frac{4}{3} \sin^2 \left(\frac{1}{3} \sin^{-1}\frac{3\; \sqrt{\frac{\Omega_d}{d} (N+2)\bar{\lambda}_k}}{2\sqrt{2}} \right).
\end{equation}

Now we substitute the values of $\bar{m}_*^2$ and $\bar{u}_*$ into the flow equation for $\bar{\lambda}_k$. This gives the next-order correction to the flow of $\bar{\lambda}_k$, which is, up to third order in $\bar{\lambda}_k$,
\begin{align}
 k \partial_k \bar{\lambda}_k &=  (d-4)\bar{\lambda}_k  + \frac{\Omega_d}{d} \bar{\lambda}_k^2  \frac{2 (N+8)}{3} 
 \nonumber\\
 &\phantom{m}
+ \frac{\Omega_d^2}{d^2} \bar{\lambda}_k^3  \frac{4(N^2 + 15N + 38) - d(N+2)(N+8) } { 3(d-3)} \nonumber\\
&\phantom{m}+ \mathcal{O}(\bar{\lambda}_k^4)
.\label{eq:improved lambda flow}
\end{align}
 In $d=4$ we find 
 \begin{align}
 k \partial_k \bar{\lambda}_k &= \bar{\lambda}_k^2 \frac{ (N+8)}{48} -  \bar{\lambda}_k^3  \frac{ (5N + 22)}{756\pi^2} + \mathcal{O}(\bar{\lambda}_k^4 ). \label{eq:correct beta term lambda}
\end{align}
This agrees with Eq.(4.9) in Ref.~\cite{PapenbrockWetterich95}, up to geometrical factors which come from using a different regulator.

There is a factor of $d-3$ in the denominator for the term $\sim \bar{\lambda}_k^3$ in Eq.~\eqref{eq:improved lambda flow}. This would indicate that the result is invalid for $d=3$. This is, however, an artifact of the expansion. We can evaluate the flow equation for $\bar{u}_k$ in $d=3$ directly to find
\begin{equation}\label{eq:neutral one field u solution}
 \bar{u}_*  (d=3) = \bar{\lambda}_k^2 \frac{N+26}{9(N+14) (1+\bar{m}_k^2)}.
\end{equation}
This gives the beta function for $\bar{\lambda}_k$:
\begin{align}
k \partial_k \bar{\lambda}_k &=  - \bar{\lambda}_k  + \frac{\Omega_3}{9} \bar{\lambda}_k^2  \frac{N^2 + 14N +120}{(N+14)(1+\bar{m}_k^2 )^3}.
\end{align}

Solving these equations in $2<d<4$, we find the usual fixed points: the trivial, Gaussian fixed point at $\bar{m}^2_* = \bar{\lambda}_* = 0$ and the nontrivial Wilson-Fisher fixed point with  $\bar{m}^2_* <0$ and $\bar{\lambda}_* > 0$. For instance, for $N=2$ and $d=3$ we find that
\begin{align}\label{eq:WF fixed point}
 \bar{m}^2_{*,\mathrm{WF}} &= -0.166, &
 \bar{\lambda}_{*,\mathrm{WF}} &= \frac{0.551}{\Omega_3}.
\end{align}

This concludes our treatment of the simple $O(N)$ model. 

\section{Neutral two-field model}\label{sec:Neutral two-field model}
We now turn our first main interest: the neutral two-field model without coupling to gauge fields. This model is a special case of the $O(N_1) \oplus O(N_2)$ models.
The most general action involving an $N_1$-component field $\phi_a$ and an $N_2$-component field $\chi_a$ with density-density coupling reads 
 \begin{align}\label{eq:neutral two-field action}
  S
  &= \int \td^d x \; \Big(
  \frac{1}{2} (\nabla \phi_a)^2 + \frac{1}{2} m_\phi^2 \phi_a^2 + \frac{1}{4!}\lambda_\phi (\phi_a^2)^2 \nonumber\\  
  &\phantom{mm} +  \frac{1}{2} (\nabla \chi_a)^2 + \frac{1}{2} m_\chi^2 \chi_a^2 + \frac{1}{4!}\lambda_\chi (\chi_a^2)^2
  + g \phi_a^2  \chi_b^2.
    \Big).
 \end{align}
 This is just two copies of the $O(N)$ model, with the added interspecies coupling proportional to $g$. For $g =12\lambda$, this model has a larger symmetry group $O(N_1 + N_2)$.
 
 It is not difficult to derive the flow equations in this most general case. Solving them, however, is complicated. Since our main interest motivated in the Introduction lies in two two-component fields, we shall immediately specialize to the case where $N_1 = N_2 =N$, and, furthermore, to the symmetric model where $m^2_\phi = m^2_\chi \equiv m^2$ and $\lambda_\phi= \lambda_\chi \equiv \lambda$. In the case in which $\lambda_\phi \neq \lambda_\chi$, it is known that this model has six fixed points near $d=4$~\cite{Kosterlitz76,Calabrese03}; by equating $\lambda_\phi= \lambda_\chi$, this number reduces to four, as we shall verify in this section. In $d=3$ we reproduce the four fixed points found in Refs.~\cite{Ceccarelli15,Ceccarelli16}. For FRG treatment of the more general model, see Ref.~\cite{Eichhorn13}.
 
 It is straightforward to obtain the fixed points of the simultaneous flow of $m^2$, $\lambda$, and $g$ of this symmetric $O(N) \oplus O(N)$ model in the FRG. However, we find that the fixed-point structure does not survive for $N=2$, which is our main case of interest, or for $N=4$. In $N=2$ and $N=4$ there are only two, instead of three, nontrivial fixed points. This might be related to the fact that, in the one-loop $\epsilon$-expansion, two fixed points merge for $N=2$~\cite{Herbut07}.  This is the reason that we use the method outlined in Sec.~\ref{sec:Corrections to the O(N)-model}; the improved flow equations for $\lambda$ and $g$ will lead to the correct structure of fixed points all the way down to $N=2$. 
 
 \subsection{Flow equations}
   We are, therefore, considering the effective action with sixth-order terms:
 \begin{widetext}
 \begin{align}\label{eq:third-order two field effective action}
  \Gamma_k = \int \td^d x \; \Big( 
  -\frac{1}{2} \nabla^2 (\phi_a^2 + \chi_a^2) & + \frac{1}{2} m_k^2 (\phi_a^2 +\chi_a^2) + \frac{1}{4!}\lambda_k \left( (\phi_a^2)^2 + (\chi_a^2)^2 \right) + g_k \phi_a^2  \chi_b^2
  \nonumber\\
   &+ \frac{1}{48} u_k \left( (\phi_a^2)^3 + (\chi_a^2)^3 \right) + \frac{1}{24} w_k \left(  (\phi_a^2)^2 \chi_b^2  +  \phi_a^2 (\chi_b^2)^2 \right)
  \Big).
 \end{align}
  \end{widetext}
  Again, we have chosen the case, which is symmetric between exchange of $\phi_a$ and $\chi_a$.
  
 We now proceed as in Sec.~\ref{sec:Corrections to the O(N)-model}. The Wetterich equation, Eq.~\eqref{eq:Wetterich equation}, is evaluated in a constant background where $\phi_a(x) = \phi_{0a} =  \delta_{a1} \phi_0$ and $\chi_a(x) = \chi_{0a} =  \delta_{a1} \chi_0$.
  Using the regulator, Eq.~\eqref{eq:Litim regulator}, the matrix contribution on the right-hand side of the Wetterich equation is of the form
  \begin{equation}\label{eq:two-field propator}
  \left(\Gamma^{(2)}_{k} + R_{k}\right)(-q,q) = \delta(0) \left( k^2 \mathbb{I}_{2N} + \mathcal{M}^2 \right)
 \end{equation}
 where $\mathbb{I}$ is the identity matrix and $\mathcal{M}^2$ is the ``mass matrix,''  given explicitly in Appendix~\ref{subsec:appendix neutral two field}, with eigenvalues $\mathcal{M}^2_{(i)}$.  These are independent of $q$, and the integral over $q$ can be performed trivially:
\begin{align}\label{eq:two-field flow equation integrated}
 k\partial_k \Gamma_k 
 &= \frac{k}{2} \int \td^d q\; 2k \Theta(k^2 - q^2) \sum_i  \frac{1}{k^2 + \mathcal{M}^2_{(i)} }\nonumber\\
 &= \frac{\Omega_d k^{d+2}}{d} \sum_i  \frac{1}{k^2 + \mathcal{M}^2_{(i)} }.
\end{align}
 We can now carry out the sum on the right-hand side and perform a series expansion in the fields. Comparing term by term with the left-hand side of Eq.~\eqref{eq:Wetterich equation} leads to the flows of the couplings. We are interested in the flow of dimensionless parameters, so we rescale $\bar{m}^2_k = k^{-2} m_k^2$, $\bar{\lambda}_k = k^{d-4} \lambda_k$, $\bar{g}_k = k^{d-4} g_k$, $\bar{u}_k = k^{2d-6} u_k$, $\bar{w}_k = k^{2d-6} w_k$. The flow equations of the couplings are:
\begin{widetext}
\begin{align}
  k \partial_k \bar{m}^2_k  
 &= -2\bar{m}^2_k  - \frac{\Omega_d}{d} \frac{(N+2)\bar{\lambda}_k  + 12\bar{g}_k N}{3(1 + \bar{m}^2_k )^2},\label{eq:flow equation m}\\
 k \partial_k \bar{\lambda}_k 
 &=  (d-4)\bar{\lambda}_k + \frac{\Omega_d}{d} \frac{2(N+8) \bar{\lambda}_k^2 + 288 N \bar{g}_k^2 - (1 + \bar{m}^2_k ) \Big(9 (N+4)\bar{u}_k + 6 N \bar{w}_k\Big)}{3 (1 + \bar{m}^2_k )^3},\label{eq:flow equation lambda}\\
 k \partial_k \bar{g}_k
 &=   (d-4)\bar{g}_k +\frac{\Omega_d}{d} \frac{96\bar{g}_k^2 + 4 (N+2) \bar{\lambda}_k \bar{g}_k - (1 + \bar{m}^2_k ) (N+2) \bar{w}_k}{3 (1 + \bar{m}^2_k )^3},\label{eq:flow equation g}\\
k \partial_k \bar{u}_k
 &=  (2d-6)\bar{u}_k + \frac{\Omega_d}{d} \frac{2}{9 (1 + \bar{m}^2_k )^4} \Big[-(N+26) \bar{\lambda}_k^3 - 1728N \bar{g}_k^3 
 + (1 + \bar{m}^2_k )\Big(9(N + 14) \bar{\lambda}_k \bar{u}_k + 72 N \bar{g}_k \bar{w}_k \Big)\Big] , \label{eq:flow equation u} \\
k \partial_k \bar{w}_k
 &= (2d-6)\bar{w}_k + \frac{\Omega_d}{d} \frac{2}{ (1 + \bar{m}^2_k )^4} \Big[ -1152 \bar{g}_k^3 - 24 (N+14) \bar{g}_k^2\bar{\lambda}_k - 2 (N+8) \bar{g}_k \bar{\lambda}_k^2 \nonumber\\
 &\phantom{mmmmmmmmmmmmmmmmmm} + (1 + \bar{m}^2_k )\Big( (N + 6) \bar{\lambda}_k \bar{w}_k + 6 (N+4) \bar{g}_k \bar{u}_k + 8(N+10) \bar{g}_k \bar{w}_k \Big)\Big].\label{eq:flow equation w}
\end{align}
\end{widetext}
These reduce to the one-field flows, Eqs.~\eqref{eq:one-field flow m}--\eqref{eq:one-field flow u}, when taking $g=w=0$.
There are a few things to note immediately about these flow equations. First, the flow of $\bar{m}^2_k $ is not affected by the higher-order terms. Second, as expected, the flows of $\bar{\lambda}_k$ and $\bar{g}_k$ are affected. As before, the strategy is now to solve the equations for $\bar{u}_k$ and $\bar{w}_k$ first, and put those values in the flow equations for $\bar{\lambda}_k$ and $\bar{g}_k$. Third, we see that, just as $d=4$ is special for the flow of $\bar{\lambda}_k$ and $\bar{g}_k$ because the first term on the right-hand side disappears, $d=3$ is special for the flows of $\bar{u}_k$ and $\bar{w}_k$. Since $d=3$ is our main case of interest, and since it is computationally simpler, we will present results in $d=3$ only. Some comments on other dimensions can be found at the end of this section.

\subsection{Fixed points}
We now derive the fixed points of Eqs.~\eqref{eq:flow equation m}--\eqref{eq:flow equation w} in $d=3$.
First, we solve the flow equation for $\bar{m}_k^2$. We use Eq.~\eqref{eq:flow m general solution} with $\alpha = \frac{\Omega_d}{6d} \big[(N+2)\bar{\lambda}_k + 12 N \bar{g}_k\big]$ to find
\begin{equation}
 \bar{m}^2_{*} = -\frac{4}{3} \sin^2 \left(\frac{1}{3} \sin^{-1}\frac{3\; \sqrt{\frac{\Omega_d}{d} \big[(N+2)\bar{\lambda}_k + 12 N \bar{g}_k\big]}}{2\sqrt{2}} \right).
\end{equation}

The first terms on the right-hand sides of Eqs.~\eqref{eq:flow equation u} and \eqref{eq:flow equation w} vanish in $d=3$. As a simplification the denominators $(1+\bar{m}^2_k)^4$, which must be assumed to be nonvanishing, can be factored out. We find two independent equations, with solutions
\begin{widetext}
\begin{align}
\bar{u}_* &= -\frac{13824(\bar{g}_k^3 N \big(\bar{\lambda}_k - (N+4)\bar{g}_k\big) + 144 \bar{g}_k^2 \bar{\lambda}_k^2 N (N+8) - \bar{\lambda}_k^3 (N+26) \big( (N+6)\bar{\lambda}_k + 8 (N+10) \bar{g}_k \big)}{9 (1+\bar{m}^2) \Big(48 \bar{g}_k^2 N (N+4)-8 \bar{g}_k \bar{\lambda}_k \left(N^2+24 N+140\right)-\bar{\lambda}_k^2 \left(N^2+20 N+84\right)\Big)},
\label{eq:u fixed point}
\\
\bar{w}_* &= -\frac{4\bar{g}_k \Big(864 \bar{g}_k^2 \big((N+14)\bar{\lambda}_k - N(N+4)\bar{g}_k \big) + \bar{\lambda}_k^2 \big( (N^2+18N+116)\bar{\lambda}_k + 18(N+14)^2 \bar{g}_k \big) \Big)}{3 (1+\bar{m}^2) \Big(48 \bar{g}_k^2 N (N+4)-8 \bar{g}_k \bar{\lambda}_k \left(N^2+24 N+140\right)-\bar{\lambda}_k^2 \left(N^2+20 N+84\right)\Big)}.
\label{eq:w fixed point}
\end{align}
\end{widetext}

We now substitute these solutions, $\bar{m}^2_{*}$, $\bar{u}_*$, $\bar{w}_*$ as functions of $\bar{\lambda}$ and $\bar{g}$, into the flow equations for $\bar{\lambda}$ and $\bar{g}$, i.e., Eqs.~\eqref{eq:flow equation lambda} and \eqref{eq:flow equation g}. The expressions are too long to write down explicitly. The flow equations $k \partial_k \bar{\lambda}_k = 0$ and $k \partial_k \bar{g}_k = 0$ are solved simultaneously; we do this numerically for certain values of $N$. We emphasize again that it is necessary to include the higher-order corrections to the flow equations to obtain all four fixed points in $N=2$ and $N=4$.
 
\begin{figure}
  \includegraphics[width=.45\textwidth]{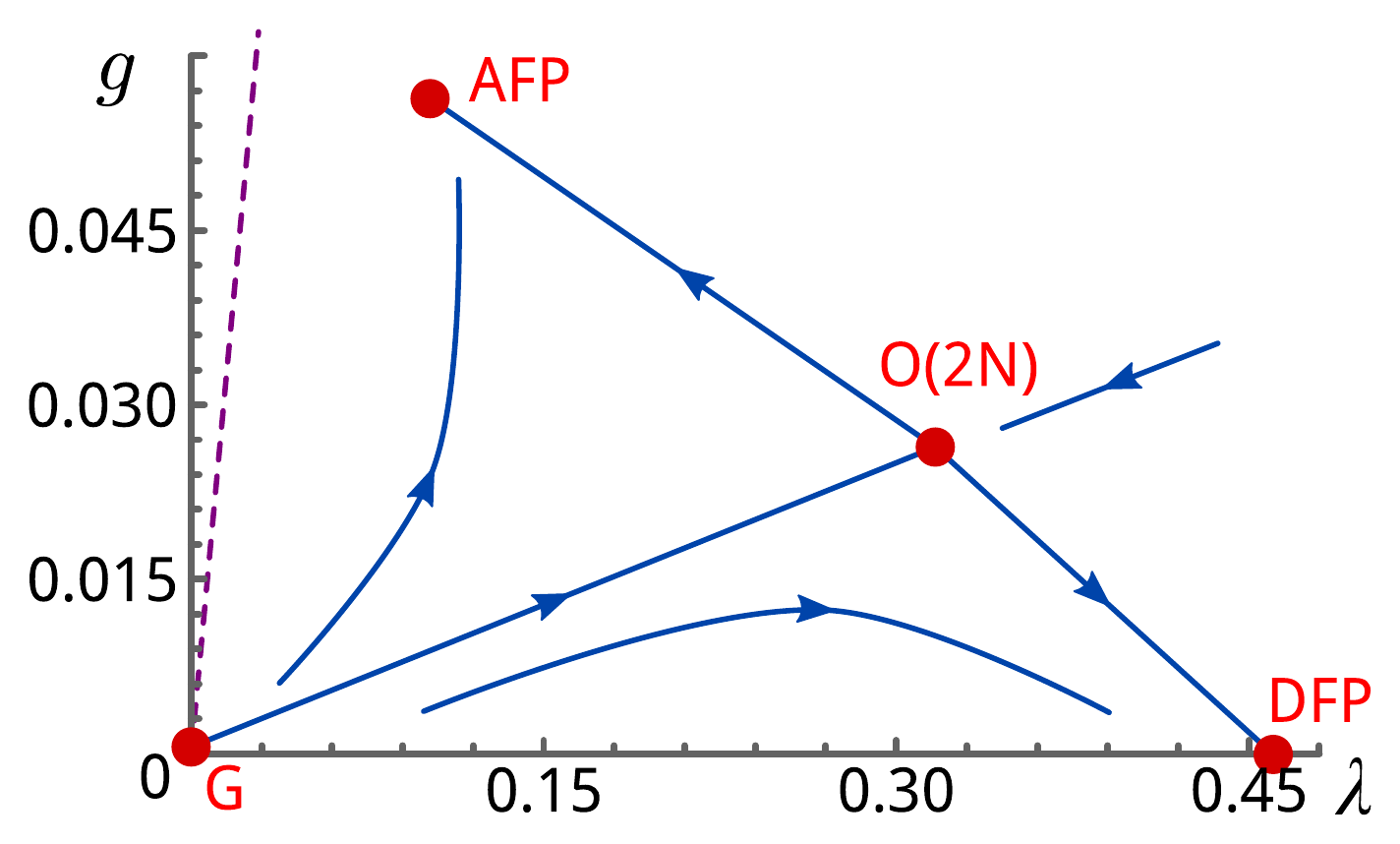}
  \caption{Fixed points of the neutral two-field model, Eq.~\eqref{eq:neutral two-field action}, in $d=3$ for $N=3$. Values of the fixed points (in units of $\Omega_3 = 1/2\pi^2$) correspond to those in Table~\ref{table:neutral two-field fixed points}. The dashed purple line indicates where the denominator of $u_*$ and $w_*$ in Eqs.~\eqref{eq:u fixed point} and ~\eqref{eq:w fixed point} vanishes, signaling that the approximation is no longer valid in the region left of this line.}\label{fig:neutral two-field flow diagram}
 \end{figure}

We find these fixed points for all $N \ge 2$ in $d=3$. We follow the nomenclature of Refs.~\cite{Ceccarelli15,Ceccarelli16}. Numerical values for $N=3$ are given in Table~\ref{table:neutral two-field fixed points}, and the flow diagram for $N=3$ is pictured in Fig.~\ref{fig:neutral two-field flow diagram}. 
 The trivial fixed point (``G'' for Gaussian) has $\bar{\lambda}_* = \bar{g}_* = 0$ and is unstable. There is one fixed point where $\bar{\lambda}_* = 12\bar{g}_*$, and this has therefore enhanced the symmetry of $O(2N)$; see Eq.~\eqref{eq:neutral two-field action}. This fixed point has one stable and one unstable direction and separates between the two regions in parameter space where either one or two fields condense. At the infrared stable decoupled fixed point (DFP) the interspecies coupling flows to zero, $\bar{g}_* = 0$, and the two fields behave as independent $O(N)$ fields with $\bar{\lambda}_*$ at their WF fixed point; cf. Eq.~\eqref{eq:WF fixed point}. Here two fields condense, and because of the symmetry of the model, they do so in the same way. It corresponds to the miscible phase in two-component BECs. The asymmetric fixed point (AFP) is also infrared stable in the $\bar{\lambda}_k$-$\bar{g}_k$ space. It has a repulsive coupling $\bar{g}_*$ large enough to prevent both fields from condensing at the same time, and it corresponds to the immiscible phase in two-component BECs.  Note than when the denominator in  Eqs.~\eqref{eq:u fixed point} and ~\eqref{eq:w fixed point} vanishes, the third-order couplings $u_*$, $w_*$ change sign. This indicates that the approximation we are using is no longer valid. The fixed points that we find below are well within the region of validity, as indicated in Fig.~\ref{fig:neutral two-field flow diagram}.
 
We note that, for $N=2$, we numerically find another fixed point, very close to the DFP with $g$ slightly negative. At present we cannot determine whether this is an artifact of the approximation. We leave this for future investigation.

\begin{table}
\hfill
\begin{tabular}{lcc}
& $\bar{\lambda}_* \Omega_3$  & $\bar{g}_* \Omega_3$ \\
 \hline
G &  0 & 0 \\
DFP&   0.46 &  0  \\
$O(2N)$ &  0.32 & 0.026 \\
AFP &  0.10 & 0.056
\end{tabular}
\hfill\null

\caption{Fixed points of the neutral two-field model Eq.~\eqref{eq:neutral two-field action}, in $d=3$ for $N=3$.}\label{table:neutral two-field fixed points}
\end{table}

The FRG derivation is valid for all $2<d<4$, even though it is numerically more difficult if $d \neq 3$. We have looked at the evolution of the fixed-point structure as $d$ is increased from $d=3$ to $d=3.99$. The decoupled and $O(2N)$ fixed points are robust and exist in this entire range. However, the asymmetric fixed point moves to the region where $\bar{\lambda}_* < 0$ around $d=3.6$ for $N=3$, and it must be rejected since negative self-coupling leads to an instability. At around $d=3.8$ the DFP splits into two, and we get an additional fixed point with $\bar{\lambda}_* >0$ and $\bar{g}_* >0$. It is known that in the symmetric two-field model the $\epsilon$-expansion also leads to such a fixed point, called the {\em biconical} fixed point~\cite{Kosterlitz76}, and that it is distinct from the AFP~\cite{Ceccarelli16}. This is consistent with our findings.

\section{Charged two-field model}\label{sec:Charged two-field model}
Now we will turn to the charged model including the coupling to the gauge fields. Compared to the neutral case, the FRG treatment is severely complicated by the presence of a gauge freedom. Introducing a momentum scale $k$ typically violates gauge invariance, so the gauge must be fixed very carefully to assure that the final results are consistent~\cite{FejosHatsuda16,FejosHatsuda17}.
We shall first review the method developed in Refs.~\cite{FejosHatsuda16,FejosHatsuda17} for the one-field model, before tackling the two-field model. In this section we will use $M = N/2$ for the number of complex components. 

\subsection{One-field Abelian-Higgs model}\label{subsec:One-field Abelian-Higgs model}
We shortly review the one-field Abelian-Higgs model, and outline how the results are affected by adding a sixth-order potential term as in Sec.~\ref{sec:Corrections to the O(N)-model}.

Consider an $M$-component complex field $\Phi_a = (\sigma_a + \mathrm{i} \pi_a)/\sqrt{2}$ where $\sigma_a$ and $\pi_a$ are real fields, minimally coupled to a dynamic $U(1)$ vector-gauge field $A_i$ with the action
\begin{align}\label{eq:AH action}
 S_\mathrm{AH} &= \int \td^d x \Big( \frac{1}{2} (\partial_i A_j - \partial_j A_i)^2 +  \lvert (\partial_i + \ti e A_i) \Phi_a\rvert^2\nonumber\\
 &\phantom{mmmmm} + m^2 \lvert \Phi_a \rvert^2 + \frac{\lambda}{6} (\lvert \Phi_a \rvert^2 )^2 \Big).
\end{align}
Here $e$ is the (electric) charge, denoting the strength of the coupling between the gauge field and the scalar field. At $M=1$ this is equivalent to the Ginzburg-Landau model of superconductivity. 

Now we discuss the method used in Refs.~\cite{FejosHatsuda16,FejosHatsuda17}. First of all, the action, Eq.~\eqref{eq:AH action}, has a gauge freedom that must be taken care of. This is done by adding a gauge-fixing term,
\begin{equation}\label{eq:gauge fixing Lagrangian AH}
 \mathcal{L}_\mathrm{gf} =  \frac{1}{2\xi} (\partial_i A_i + \xi e \tilde{\sigma}_a \pi_a)^2.
\end{equation}
Here $\tilde{\sigma}_a$ is a freely chosen nondynamical dummy field degree of freedom, and $0 \le \xi<\infty$ is the gauge-fixing parameter. This corresponds to a $R_\xi$-gauge-fixing function $G = \frac{1}{\sqrt{\xi}} ( \partial_i A_i + \xi e \tilde{\sigma}_a \pi_a)$, and it can be implemented by introducing a by pair of Grassmann-valued ghost fields $c^*, c$ via
\begin{equation}\label{eq:ghost Lagrangian}
 \mathcal{L}_\mathrm{ghost} = c^{*} \left(- \partial^2 + \xi e^2 \tilde{\sigma}_a \sigma_a \right) c.
\end{equation}

Second,  now it is necessary to include wave function renormalization factors and charge rescaling as follows:
\begin{align}\label{eq:field rescaling}
 \Phi_a &\to \sqrt{Z_{k}}\Phi_a, &
 A_i &\to \sqrt{Z_{A,k}} A_i, &
 e &\to  \frac{Z_{e,k}}{\sqrt{Z_{A,k}}Z_{k}} e.
\end{align}
We assume that the Ward identity is not violated, so $Z_{e,k} = Z_{k}$, which also implies that the rescaling of the charge is consistent with gauge invariance of the covariant derivative.
The full effective action in the local potential approximation reads:
\begin{align}\label{eq:AH effective action}
 \Gamma_k &= \int \td^d x \Big[ 
 \frac{1}{2}Z_{A,k}A_i \left( - \partial^2 \delta_{ij} + \partial_i \partial_j(1-\xi^{-1}) \right) A_j 
 \nonumber\\
 &\phantom{m}+  \frac{1}{2}Z_k \sigma_a\left(-\partial^2 + m_k^2  + \frac{1}{12}Z_k\lambda_k \sigma_b^{2} + e^2 A_i A_i \right)\sigma_a 
 \nonumber\\
 &\phantom{m}
  +  \frac{1}{2}Z_k \pi_a \left(-\partial^2+  m_k^2  + \frac{1}{12}Z_k\lambda_k \pi_b^{ 2} + e^2 A_i A_i \right)\pi_a
 \nonumber\\
 &\phantom{m}
 + \frac{1}{12}Z_k^2 \lambda_k \sigma_a^{ 2} \pi_b^{ 2}  + \xi_k  \frac{Z_k}{Z_{A,k}} e^2 \pi_a \tilde{\sigma}_a\tilde{\sigma}_b \pi_b +  
 \nonumber\\
 &\phantom{m} -   Z_k e (\partial_i A_i) (\sigma_a - \tilde{\sigma}_a)  \pi_a  -  2Z_k e A_i\pi_a \partial_i \sigma_a
 \nonumber\\
 &\phantom{m}
  + c^{*} \left(- \partial^2 + \xi e^2 \frac{Z_k}{Z_A} \tilde{\sigma}_a \sigma_a \right) c \Big] .
\end{align}
Note that we allow the gauge fixing parameter to flow, $\xi = \xi_k$.

As it turns out, terms with finite mass $m_k^2$ lead to ambiguities in calculating the flow of $Z_{A,k}$. For that reason, it is going to be assumed that $m_k^2$ flows to zero at the fixed points as $k\to0$, and all expressions will be evaluated at $m_k^2 =0$~\cite{FejosHatsuda16,FejosHatsuda17}. 
Similarly, the flow equation generates a mass term for the gauge field $m_{A,k}^2$, which breaks gauge invariance, but this term identically flows to zero as $k \to 0$ once it is properly adjusted at the UV scale, and it can therefore be safely set to zero through all calculations.

As before, we will evaluate the left- and right-hand sides of the Wetterich equation, Eq.~\eqref{eq:Wetterich equation}, to determine flow equations. We will use the following terms on the left-hand side:
\begin{align}
 & \partial_k ( Z_k^2 \lambda_k) (\sigma_a^2)^2, \qquad 
 (\partial_k Z_k) \sigma_a (-\partial^2) \sigma_a, \nonumber\\
 &(\partial_k Z_{A,k}) A_i \left( - \partial^2 \delta_{ij} + \partial_i \partial_j (1-\xi_k^{-1}) \right) A_j.
\end{align}
Therefore, in evaluating the flow for $Z_k \lambda_k^2$, we can assume a background where all fields are vanishing except for a constant $\sigma_a = \sigma_{a0}$. For the flow of $Z_k$ itself, owing to the gauge fixing and the choice $\tilde{\sigma}_a  = \sigma_{a0}$, we can also treat $\partial_i \sigma_a$ as a constant vector field, while all other fields vanish. Using this flow equation we can determine the flow of the coupling constant using
\begin{align}\label{eq:Zk substituted flow}
  \partial_k \lambda_k &= \frac{1}{Z^2_k}\left( \partial_k (Z^2_k \lambda_k) - \lambda_k 2 Z_k (\partial_k Z_k)\right).
\end{align}
The flow of $Z_{A,k}$ is evaluated in a background where only $A_i$ is nonvanishing. The flow of the charge follows from that of $Z_{A,k}$ via the definition $e_k^2 = e^2 / Z_{A,k}$.

For evaluating the right-hand side of the Wetterich equation Eq.~\eqref{eq:Wetterich equation} we need to choose the regulator matrix. This is done as follows: First, the propagator matrix $\Gamma^{(2)}_k$ is calculated and diagonalized. Then, to each eigenvalue $\gamma^{(i)}_k$ thus obtained, a regulator $R^{(i)}_k(q) = Z^{(i)}_k R_k(q)$ where $R_k(q)$ given in  Eq.~\eqref{eq:Litim regulator}, is added. The factor $Z^{(i)}_k$ depends on the prefactor of the $q^2$ term that is obtained in each $\gamma^{(i)}_k$. By neglecting the anomalous dimensions (which produce only higher-order terms) the right-hand side is then of the form
\begin{equation}\label{eq:AH Wetterich equation}
 \mathrm{rhs} = \frac{1}{2} \sum_i \int \td^d q\; \frac{2k Z^{(i)}_k}{\gamma^{(i)}_k  + R^{(i)}_k(q)} \Theta(k^2 - q^2).
\end{equation}
This is then evaluated explicitly, expanded in a series in the fields and compared with the left-hand side.

We refer the reader to Ref.~\cite{FejosHatsuda17} for further details of the derivation and only state some of the results. The flow of the charge is independent of the other coupling constants; the dimensionless charge $\bar{e}^2_k = k^{d-4} e^2_k$ has two fixed points, one at $\bar{e}^2_k = 0$ and the other at
\begin{align}\label{eq:charge fixed points}
  \bar{e}_*^2 &= \frac{d(d+2)(4-d)}{8M \Omega_d}.
\end{align}
Even though at the level of the effective action Eq.~\eqref{eq:AH effective action} we are free to choose any value of the gauge-fixing parameter $\xi_k$, it turns out that, in the present FRG scheme, there is only one choice consistent with the flow equation:
\begin{equation}\label{eq:xi consistent}
 \xi_k = \frac{2}{4-d}.
\end{equation}

The flow of $Z_k$ depends only on the charge $e^2_k$:
\begin{equation}\label{eq:Zk flow}
 k \partial_k Z_k = \frac{8}{d-2} \frac{\Omega_d}{d} k^{d-4} (d-1 + \xi_k) e_k^2 Z_k.
\end{equation}
With all of this, it is found that the fixed points are those in Fig.~\ref{fig:AH flow diagram} {\em for all} $M$.

We now wish to comment on introducing the term
$\frac{u}{6}(\Phi_a\Phi_a)^3$ into the ansatz of $\Gamma_k$, as we did in
Sec.~\ref{sec:Corrections to the O(N)-model}. We follow the procedure exactly as before: we use the term
$\partial_k (Z_k^3 u_k)(\sigma_a^2)^3$ to determine the flow of $u_k$,
which is then solved in its fixed point, and the solution for $u_k$ is
substituted back into the flow of $\lambda_k$. For the neutral fixed points,
this gives the same structure as before, but note that, since we are
working with $m^2_k = 0$, results differ quantitatively. It is
important to mention, however, that, in the charged case, the introduction
of $u_k$ makes the tricritical fixed point disappear. The reason is that
when substituting the fixed point expression for $u_k$ into the flow of
$\lambda_k$, a new fixed point for $\lambda_k$ may arise. If the charge is
zero, then this is identical to the trivial fixed point, but for $e^2_k \neq
0$, all roots for $\lambda_k$ are different. At $e^2_k = e^2_*$, two of them
become complex, leaving the superconducting fixed point C as the only real
solution (see Fig.\ref{fig:AH flow diagram}). As pointed out already, the introduction of $u_k$ is necessary
for reproducing the fixed-point structure of the two-field model, and 
therefore we do not wish to discard it. The disappearance of the tricritical fixed-point could be related to the fact that, for small $\lambda_k$, the fixed point value for $u_k$ gets close to zero, leading to a singularity in the coupled flow equations of $\lambda_k$ and $g_k$. It would be interesting to see effects of even higher order contributions, but we leave that for further studies.

\subsection{Two-field effective action}
We now turn to the two-field model coupled to two gauge fields. We are considering two complex-valued $M$-component fields $\Phi^f_a$ and two vector gauge fields $A^f_i$, labeled by three indices: $i = 1,\ldots,d$ as the spacetime index; $a =1,\ldots,M$ as the vector index, and $f=1,2$ as the flavor index.  

The total action is
\begin{widetext}
\begin{align}\label{eq:2field AH Lagrangian}
 S &= \int \td^d x \Big[ \sum_{f=1,2} \Big( \frac{1}{2} A^f_i \left( - \partial^2 \delta_{ij} + \partial_i \partial_j(1-\xi^{-1}) \right) A^f_j+(D^f_i \Phi^f_a)^\dagger D^f_i \Phi_a^f  + m^2 \Phi_a^{f\dagger} \Phi^f_a + \frac{\lambda}{6} (\Phi_a^{f\dagger} \Phi^f_a)^2+ \frac{u}{6}(\Phi_a^{f\dagger} \Phi^f_a)^3 \Big) \nonumber\\
 &\phantom{mmmmm}+ 4g (\Phi_a^{1\dagger} \Phi^1_a)(\Phi_b^{2\dagger} \Phi^2_b)  + \frac{w}{3} \big( (\Phi_a^{1\dagger} \Phi^1_a)^2(\Phi_b^{2\dagger} \Phi^2_b) +  (\Phi_a^{1\dagger} \Phi^1_a)(\Phi_b^{2\dagger} \Phi^2_b)^2 \big) \Big].
\end{align}
\end{widetext}
Here $D^f_i \Phi^f_a = (\partial_i + \ti e^f A^f_i) \Phi^f_a$, and we assumed the symmetric situation where $m^1 = m^2 =m$, $\lambda^1 = \lambda^2 = \lambda$, $e^1 = e^2 = e$, and also $M$ complex components for both fields. This model is essentially two copies of the  $M$-component Abelian-Higgs  model which are coupled only through the terms in the last line.

To avoid confusion, we will sometimes forgo the flavor indices and write instead the two fields with separate symbols, namely,
\begin{align}
 \Phi^1_a &= \tfrac{1}{\sqrt{2}} ( \sigma^1_a + \ti \pi^1_a)  \equiv  \tfrac{1}{\sqrt{2}} ( \sigma_a + \ti \pi_a),\\
 \Phi^2_a &=  \tfrac{1}{\sqrt{2}} ( \sigma^2_a + \ti \pi^2_a) \equiv \tfrac{1}{\sqrt{2}} ( \varsigma_a + \ti \varpi_a).
\end{align}
With these definitions, the coupling constants $m^2$, $\lambda$, $g$, $u$, $w$ are the same as for the uncharged model of Eq.~\eqref{eq:neutral two-field action} in terms of the real fields $\phi = (\sigma,\pi)$, $\chi = (\varsigma,\varpi)$.

This model is invariant under two separate $U(1)$-gauge transformations. We choose $R_\xi$ gauge fixes for both fields by adding two pairs of ghost fields, $c^{f*}, c^f$, $f=1,2$, with each pair governed by a Lagrangian Eq.~\eqref{eq:ghost Lagrangian}. Similarly, we carry out the rescalings Eq.~\eqref{eq:field rescaling} for both sets of fields. 

In total we find the two-field generalization of the effective action, Eq.~\eqref{eq:AH effective action}, without the higher-order terms:
\begin{align}\label{eq:2 field AH effective action}
 \Gamma_k &= \sum_{f=1,2} \Gamma^f_{k} + \int \td^d x\;  Z_k^2 g \left( \sigma_a^2 + \pi_a^2 \right)  \left( \varsigma_b^2 + \varpi_b^2 \right).
\end{align}
Here $\Gamma^f_k$ are given by Eq.~\eqref{eq:AH effective action} for the respective fields. To correctly reproduce the fixed-point structure for the neutral case for low values of $M$, as we have seen in the previous section, it is necessary to take into account the terms of sixth order in the fields in Eq.~\eqref{eq:2field AH Lagrangian}.

\subsection{Flow equations}\label{subsec:charged two-field flow equations}
We will determine the flow of the couplings at hand of the following terms in the left-hand side of the Wetterich equation Eq.~\eqref{eq:Wetterich equation}:
\begin{align}
   &\partial_k ( Z_k^2 \lambda_k) (\sigma_a^2)^2,  \quad \partial_k ( Z_k^3 u_k) (\sigma_a^2)^3, 
 \quad (\partial_k Z_k) \sigma_a (-\partial^2) \sigma_a, \nonumber\\
 & (\partial_k Z_{A,k}) A_i \left( - \partial^2 \delta_{ij} + \partial_i \partial_j \right) A_j ,
 \nonumber\\
 &\partial_k (Z_k^2 g_k) \sigma_a^2 \varsigma_b^2, \quad \frac{1}{24}\partial_k (Z_k^3 w_k) \sigma_a^2 \varsigma_b^4.
\end{align}
The addition of the third-order terms to Eq.~\eqref{eq:2 field AH effective action} will lead to new terms in the elements of the propagator matrix. 
Here we list the ones that affect the flow of the couplings, and we have already set $\pi_{a0}=\varpi_{a0} = 0$:
\begin{widetext}
\begin{align}
 \Gamma^{(2),\mathrm{3rd}}_{\sigma_a \sigma_b} &= 2Z_k^2g  \delta_{ab}\varsigma_c^2
  + \frac{1}{8}Z_k^3 u_k \sigma_c^2 \Big( \delta_{ab} \sigma_d^2 + 4\sigma_a \sigma_b  \Big)
  + \frac{1}{12}Z_k^3 w_k \varsigma_c^2  \Big( \delta_{ab}  \left( 2\sigma_d^2 + \varsigma_d^2 \right) 
  + 4\sigma_a \sigma_b\Big),\\
 \Gamma^{(2),\mathrm{3rd}}_{\pi_a \pi_b} &= 2Z_k^2g \delta_{ab} \varsigma_c^2
  + \frac{1}{8}Z_k^3 u_k \delta_{ab} \left( \sigma_c^2 \right)^2
  + \frac{1}{12}Z_k^3 w_k \varsigma_c^2   \Big( \delta_{ab}\left( 2\sigma_d^2 +\varsigma_d^2 \right) \Big),\\
 \Gamma^{(2),\mathrm{3rd}}_{\varsigma_a \varsigma_b} &=  2Z_k^2 g\delta_{ab} \sigma_c^2 
  + \frac{1}{8}Z_k^3 u_k \Big( \delta_{ab} \left( \varsigma_c^2 \right)^2 + 4\varsigma_a \varsigma_b \varsigma_c^2  \Big)
  + \frac{1}{12}Z_k^3 w_k \sigma_c^2   \Big( \delta_{ab}\left( 2\varsigma_d^2 +\sigma_d^2 \right)  + 4\varsigma_a \varsigma_b \Big),\\
\Gamma^{(2),\mathrm{3rd}}_{\varpi_a \varpi_b} &=  2Z_k^2g\delta_{ab}  \sigma_c^2
  + \frac{1}{8}Z_k^3 u_k  \delta_{ab} \left( \varsigma_c^2 \right)^2 
  + \frac{1}{12}Z_k^3 w_k \sigma_c^2  \Big( \delta_{ab} \left( 2\varsigma_d^2  +\sigma_d^2 \right)  \Big).
\end{align}
\end{widetext}
We can see that all terms involve at least one factor of $\sigma_a$ or $\varsigma_a$. Therefore, the flows of $Z_k$ and $Z_{A,k}$ are unaffected. This immediately implies that the flow of the charge is unaffected as well, and the fixed points of the charge are given by those of the one-field model Eq.~\eqref{eq:charge fixed points}.

We can now diagonalize the propagator matrix. The full expressions for the eigenvalues $\gamma_{(i)}$ are given in Appendix~\ref{subsec:appendix charged two field}. The right-hand side of the Wetterich equation is given by Eq.~\eqref{eq:AH Wetterich equation} where the regulators $R^{(i)}_k(q) = Z^{(i)}_k R_k(q)$ are chosen as before. Substituting the flow for $Z_k$ into Eq.~\eqref{eq:Zk flow}, we find the flows of the couplings using the analogues of Eq.~\eqref{eq:Zk substituted flow}. They are
\begin{widetext}
 \begin{align}
 k \partial_k \bar{\lambda}_k  &= (d-4)\bar{\lambda}_k + \frac{\Omega_d}{3d} \Big[ 2(2M+8) \bar{\lambda}_k^2 -  \frac{48d}{d-2}  \bar{e}_k^2 \bar{\lambda}_k + 72(d-1)\bar{e}_k^4 + 288(2M)\bar{g}_k^2
 - 9(2M+4)\bar{u}_k - 6(2M)\bar{w}_k \Big],\\
 k \partial_k \bar{g}_k &= (d-4)\bar{g}_k + \frac{\Omega_d}{3d} \Big[ 96 \bar{g}_k^2 + 4\bar{g}_k \bar{\lambda}_k(2+2M) -  \frac{48d}{d-2}  \bar{g}_k \bar{e}_k^2 - \bar{w}_k (2+2M) \Big],\\
 k \partial_k  \bar{u}_k &= (2d-6)\bar{u}_k +  \frac{2\Omega_d}{9d} \Big[
 - (2M + 26)\bar{\lambda}_k^3 - 1728 (2M) \bar{g}_k^3  \label{eq:charged two field flow u}
 - 18 \xi_k \bar{e}_k^2 \bar{\lambda}_k^2 -108 \xi_k^2 \bar{e}_k^4\lambda_k
- 216(d-1) \bar{e}_k^6 \nonumber\\
 &\phantom{mmmmmmmmmmm} + 9 (2M+14) \bar{\lambda}_k \bar{u}_k + 72(2M) \bar{g}_k \bar{w}_k -   \frac{108d}{d-2}  \bar{e}_k^2 \bar{u}_k
 \Big],\\
k \partial_k  \bar{w}_k  &= (2d-6)\bar{w}_k + \frac{2 \Omega_d}{d} \Big[
 -1152 \bar{g}_k^3 - 24 \bar{g}_k^2\left( (2M + 14) \bar{\lambda}_k + 6 \xi_k \bar{e}_k^2 \right) 
 -  2\bar{g}_k \left( (2M + 8) \bar{\lambda}_k^2  + 12 \xi_k  \bar{e}_k^2\bar{\lambda}_k + 36 \xi_k^2 \bar{e}_k^4\right) 
   \nonumber\\
&\phantom{mmmmmmmmmmm}  + 6(2M+4) \bar{g}_k \bar{u}_k +   \bar{w}_k \left( (2M + 6) \bar{\lambda}_k -  \frac{12d}{d-2}  \bar{e}_k^2 + 8(2M +10)\bar{g}_k  \right) \label{eq:charged two field flow w}
\Big].
\end{align}
\end{widetext}
Here we have already used $\xi_k = 2/(4-d)$ from Eq.~\eqref{eq:xi consistent} in some substitutions for notional brevity. It can be seen that these reduce both to the neutral two-field model when setting $\bar{e}_k = 0$ and $M = N/2$ and to the one-field Abelian-Higgs model when setting $\bar{g}_k = \bar{w}_k =0$.

\subsection{Fixed points}
We wish to find the fixed points of these flow equations. Since the flow of the charge is unaffected, it has two fixed points, one neutral at $\bar{e}^2_k =0$, and one charged given by Eq.~\eqref{eq:charge fixed points}. We proceed as in Sec.~\ref{sec:Neutral two-field model}: first we evaluate $k \partial_k  \bar{u}_k = 0$ and $k \partial_k  \bar{w}_k =0$, and we substitute the solutions $\bar{u}_*$ and $\bar{w}_*$ into the flow equations for $\bar{\lambda}_k$ and $\bar{g}_k$. The expressions are too long to write down explicitly.

\begin{figure}
 \includegraphics[width=.45\textwidth]{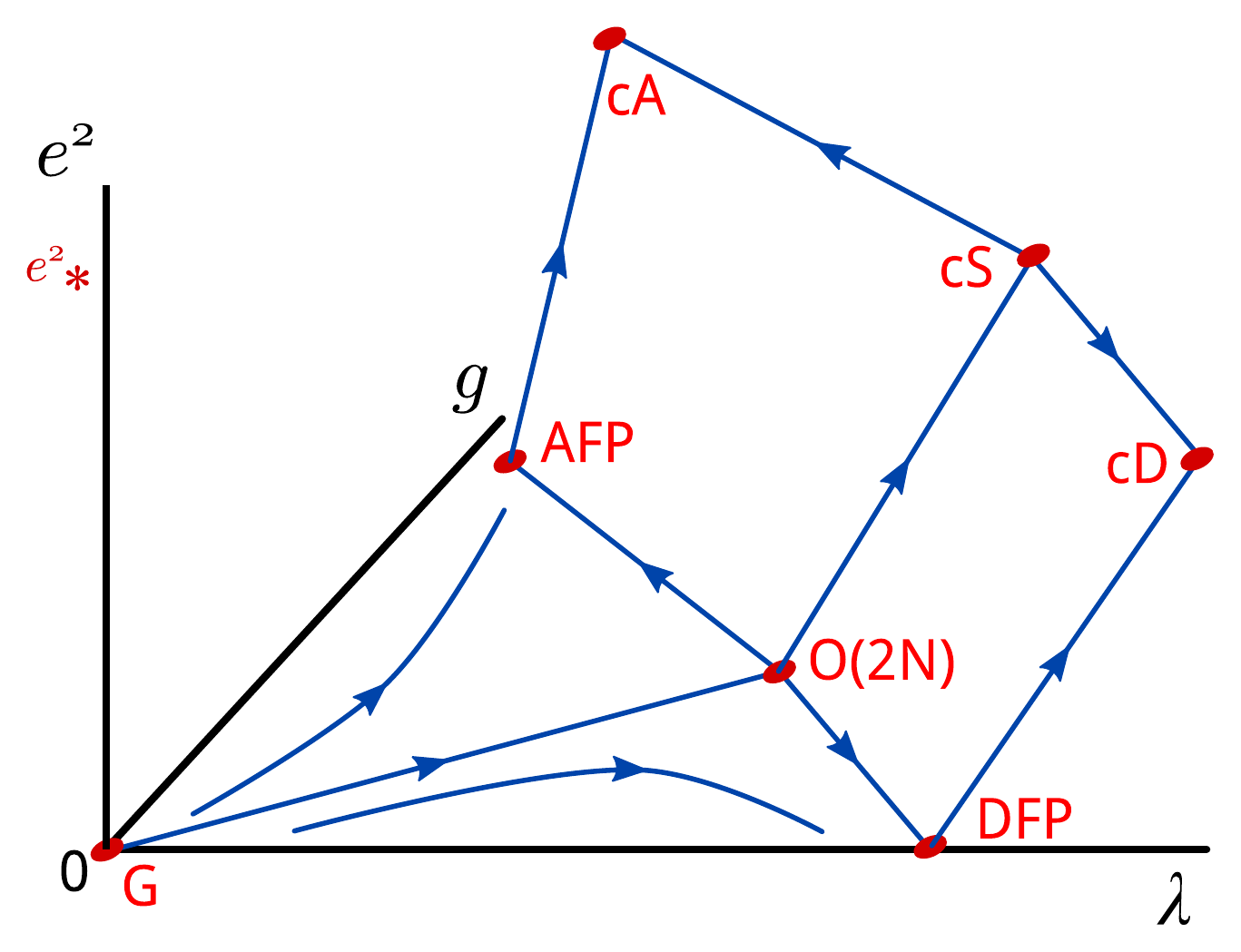}
 \caption{Sketch of the flow diagram of the charged two-field model [Eq.~\eqref{eq:2field AH Lagrangian}] for $M>1$ ($N>2$) in $d=3$, in the space of the relevant couplings $\bar{\lambda}_k$, $\bar{g}_k$, and $\bar{e}_k^2$. At $\bar{e}^2 =0$ we retrieve the four fixed points of the neutral two-field model. There are three charged fixed points cA, cS, and cD, which match the nontrivial neutral fixed points. All neutral fixed points are unstable against a finite charge.}
 \label{fig:two field charged flow diagram}
\end{figure}

We here give some results for $d=3$, which is computationally easier because the first terms on the right-hand sides of Eqs.~\eqref{eq:charged two field flow u} and \eqref{eq:charged two field flow w} drop out. For $M > 1$ the structure found in Sec.~\ref{sec:Neutral two-field model} is reproduced {\em for both the neutral and the charged fixed points}. That is, we find three nontrivial fixed points both at $\bar{e}^2_k =0$ and, more importantly, at $\bar{e}^2_k = \bar{e}^2_* =\frac{5}{2M \Omega_3}$, which are charged versions of the AFP, $O(2N)$ FP, and DFP. We therefore denote them by cA, cS, and cD. The neutral fixed points correspond of course to those  already found in the neutral model, which assures us that our approximation of setting $\bar{m}_k^2 = 0$ leads only to a small shift of the positions of the fixed points in $\bar{\lambda}_k$-$\bar{g}_k$ space, leaving the overall structure intact. As for the charged fixed points, the unstable one is no longer at the symmetric point $\bar{\lambda}_* / \bar{g}_* = 12$ because of the finite charge, but at a slightly lower ratio. The flow diagram in the parameter space of $\{ \bar{\lambda}_k, \bar{g}_k, \bar{e}^2_k \}$ is sketched in Fig.~\ref{fig:two field charged flow diagram}.
Just as we saw for the one-field Abelian-Higgs model in Sec.~\ref{subsec:One-field Abelian-Higgs model}, there is no second nontrivial fixed point at $\bar{g}_k =0$, which would correspond to the tricritical point in Fig.~\ref{fig:AH flow diagram}. As explained before, the flow equations we obtain become singular near $\bar{\lambda}_k = \bar{g}_k = 0$, which could be the reason for this absence. We view this as a point that can be improved upon in future work, whereas we focus here on the infrared stable charged fixed points.

\begin{figure}
 \includegraphics[width=.45\textwidth]{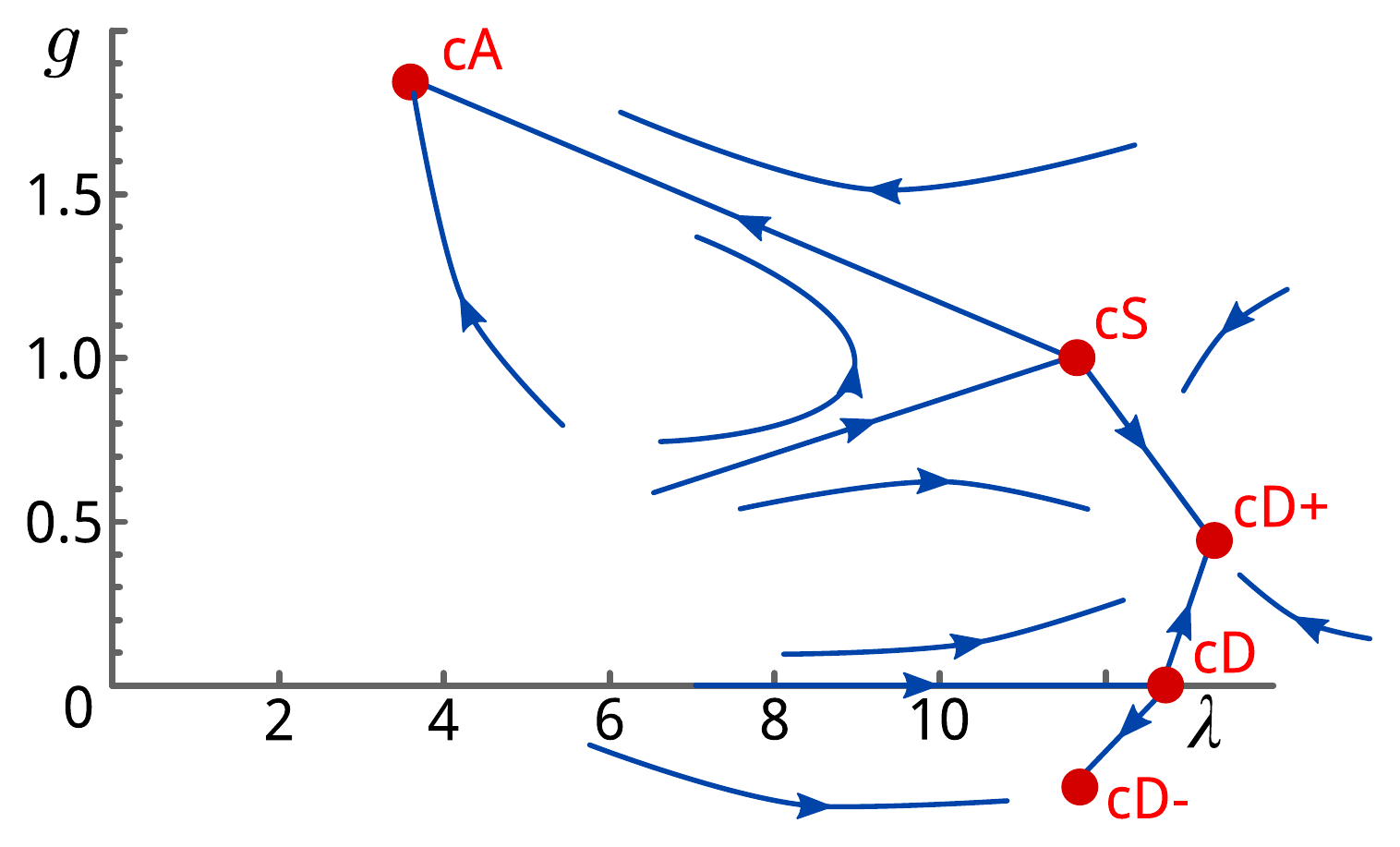}
 \caption{Sketch of the flow diagram of the charged two-field model [Eq.~\eqref{eq:2field AH Lagrangian}] for $M=1$ ($N=2$) in $d=3$, at $\bar{e}^2_* =\frac{5}{2 \Omega_3}$. The decoupled fixed point cD at $\bar{g}_* = 0$ becomes unstable toward two new infrared stable fixed points, cD$+$ and cD$-$.
 }
 \label{fig:two field charged flow diagram M1}
\end{figure}

For $M=1$, we find additional structure emerging in the flow diagram of the charged fixed points. This is sketched in Fig.~\ref{fig:two field charged flow diagram M1}. The charged decoupled fixed point cD at $\bar{g}_*=0$ becomes unstable toward two new fixed points that arise at positive and negative $\bar{g}_*$ respectively. We denote them by cD$+$ and cD$-$. Contrary to the neutral and $M>1$ charged cases, the two new fixed points are well separated from the other ones and are clearly genuine zeros of the flow equations that we have obtained.

\section{Conclusions}\label{sec:Conclusions}
We have studied the fixed points of two-scalar models with and without coupling to gauge fields. Using a minimal extension of the simplest two-field model by including the flow of the third-order coupling constants, we have been able to confirm the results of earlier works, that there are four fixed points in $d=3$ (see Fig.~\ref{fig:neutral two-field flow diagram}) for all $N \ge 2$. It may be interesting to study in more detail the fate of these fixed points as the dimension is varied, as has been done in Ref.~\cite{Borchardt16} for $2 < d \le 3$. We have confirmed the statement in Ref.~\cite{Ceccarelli16} that the asymmetric fixed point is not connected to the biconical fixed point that is found near $d=4$.

The main result of this work is that the three nontrivial fixed points of the neutral model have three siblings at finite charge; see Fig~\ref{fig:two field charged flow diagram}. This indicates that even when coupling to gauge fields, second-order phase transitions exist both to the phase where one field condenses (the infrared stable fixed point at positive $\bar{g}_*$) and to the phase where both fields condense (the infrared  stable fixed point at $\bar{g}_* = 0$). Analogous to the Abelian-Higgs models, these phase transitions take place in that part of parameter space where charge is small compared to the scalar field couplings (``large $\kappa$''). While we cannot confirm this explicitly because our flow equations become singular at small couplings, presumably in this region of the parameter space the phase transitions become first order. Interestingly, at exactly $M=1$ ($N=2$) we find the appearance of two more charged fixed points, both of them being infrared stable; see Fig.~\ref{fig:two field charged flow diagram M1}. 

While in our model the two scalar fields are coupled to two separate gauge fields, the results would be the same as when one gauge field was coupled to both scalar fields, with the same coupling constant $e$. The reason is that the flow of the charge does not depend on the coupling between the two scalar fields. In this light our results can be directly compared to, for instance, Ref.~\cite{Abreu08}.

There are some obvious extensions to carry out from here. It should not be complicated to generalize these results to less symmetric cases where $N_1 \neq N_2$ and $\lambda_\phi \neq \lambda_\chi$. It is to be expected that charged fixed points persist. Moreover, it is interesting to find out what happens to the two additional neutral fixed points that exist in these nonsymmetric cases when coupling to dynamical gauge fields.

As for the dislocation-mediated quantum melting, the existence of the charged fixed points confirms that the second-order quantum phase transition from solid to hexatic phase may exist (this would be the charged fixed point at $\bar{g}_*=0$). At present however, it is not entirely clear whether the presumed quantum phase transition would correspond to the charged or the neutral fixed point. It is known that the duality mapping employed in Refs.~\cite{Zaanen04,Beekman17} is valid only in the strongly correlated or strong type-II regime, which corresponds to a very low value of the charge $e$. For the superconducting phase transition, the duality indicates that the transition is in the Wilson-Fisher universality class, i.e., the neutral fixed point. There have been extensive discussions of the phase transition using dual methods~\cite{Kiometzis95,Hove00,Kajantie04,Nahum12}, and we cannot discern whether this has been fully resolved. We do find, however, just as for the single-field Abelian-Higgs model, that the flow is unstable against any finite charge. In any case it is interesting to study the charged fixed points in more detail, in particular by extracting critical exponents.

\begin{acknowledgments}
 We thank Sergio Benvenuti for the useful discussions.This work is supported by the Ministry of Education, Culture, Sports, Science (MEXT)-Supported Program for the Strategic Research Foundation at Private Universities “Topological Science” (Grant No. S1511006). The work of A.J.B. is also supported in part by the Japan Society for the Promotion of Science (JSPS) Grant-in-Aid for Scientific Research (KAKENHI Grant No. 18K13502). The work of G. F. is also supported by the Hungarian National Research, Development and Innovation Office (Project No. 127982).
\end{acknowledgments}

\appendix
\section{Additional calculations}\label{sec:Additonal calculations}
Here we collect some equations used in intermediate steps of several calculations.

\subsection{Neutral two-field model}\label{subsec:appendix neutral two field}
The full mass matrix $\mathcal{M}^2$ in Eq.~\eqref{eq:two-field propator} is given by
 \begin{equation}\label{eq:mass matrix}
 \mathcal{M}^2 = \begin{pmatrix}
 \mathcal{M}^2_{L,\phi} & \mathcal{M}^2_{\phi\chi} & 0 & & \cdots & &  & 0 \\
 \mathcal{M}^2_{\phi\chi} & \mathcal{M}^2_{L,\chi} & 0 & &  & & &  \\
 0            &  0         &  \mathcal{M}^2_{T,\phi} & 0 &0 &  & & \\
             &           &  0 & \ddots  & 0 & & & \vdots \\
             &           &  0 &  0 & \mathcal{M}^2_{T,\phi} & & \\
 \vdots      &           &  &&& \mathcal{M}^2_{T,\chi} & 0 & 0\\
             &           &  &&&0 & \ddots  & 0 \\
 0            &           &  & \cdots && 0 &  0 & \mathcal{M}^2_{T,\chi} \\
 \end{pmatrix}
\end{equation}
Here we have defined
\begin{widetext}
 \begin{align}
 \mathcal{M}^2_{L,\phi} &= m_{\phi,k}^2 + \frac{1}{2} \lambda_{\phi,k} \phi_0^2 + 2g_k \chi_0^2 + \frac{5}{8} u_k \phi_0^4 + \frac{1}{2} w_{k} \phi_0^2 \chi_0^2 + \frac{1}{12} w_{k} \chi_0^4 ,\\
 \mathcal{M}^2_{T,\phi} &= m_{\phi,k}^2 + \frac{1}{6} \lambda_{\phi,k} \phi_0^2 + 2g_k \chi_0^2+ \frac{1}{8} u_k \phi_0^4 + \frac{1}{6} w_{k} \phi_0^2 \chi_0^2 + \frac{1}{12} w_{k} \chi_0^4,\\
 \mathcal{M}^2_{L,\chi} &= m_{\chi,k}^2 + \frac{1}{2} \lambda_{\chi,k} \phi_0^2 + 2g_k \phi_0^2 + \frac{5}{8} u_k \chi_0^4 + \frac{1}{2} w_{k} \phi_0^2 \chi_0^2 + \frac{1}{12} w_{k} \phi_0^4,\\
 \mathcal{M}^2_{T,\chi} &= m_{\chi,k}^2 + \frac{1}{6} \lambda_{\chi,k} \phi_0^2 + 2g_k \phi_0^2+ \frac{1}{8} u_k \chi_0^4 + \frac{1}{6} w_{k} \phi_0^2 \chi_0^2 + \frac{1}{12} w_{k} \phi_0^4,\\
 \mathcal{M}^2_{\phi\chi} &= 4g_k \phi_0\chi_0 + \frac{1}{3} w_{k} \phi_0^3 \chi_0 + \frac{1}{3} w_{k} \phi_0 \chi_0^3.
\end{align}
\end{widetext}

\subsection{Charged two-field model}\label{subsec:appendix charged two field}

Here we give the eigenvalues $\gamma^{(i)}_k$ of the total propagator matrix including third-order contributions with constant fields $\sigma = \sigma_0 \neq 0$ and $\varsigma = \varsigma_0 \neq 0$, while the other fields vanish (of Sec.~\ref{subsec:charged two-field flow equations}).
The only nondiagonal contribution is from the $\sigma_1$-$\varsigma_1$ sector (the gauge field sectors can be diagonalized as in the one-field Abelian-Higgs model). This can be diagonalized to
\begin{widetext}
\begin{align}
  \gamma^{\sigma_1\varsigma_1}_{k \pm} &=  Z_k (q^2 + m_k^2) + Z_k^2 (g + \tfrac{1}{4} \lambda) (\bar{\sigma}^2 + \bar{\varsigma}^2) + \frac{1}{48}Z_k^3\left( 15 u_k (\bar{\sigma}^4 + \bar{\varsigma}^4) + 2w_k (\bar{\sigma}^4 + 12\bar{\sigma}^2\bar{\varsigma}^2 +  \bar{\varsigma}^4) \right) \nonumber\\
  &\phantom{m} \pm Z_k^2 \Big[ 
  (\tfrac{1}{4}\lambda -g)^2 (\bar{\sigma}^4 + \bar{\varsigma}^4) - 2 \left( (\tfrac{1}{4}\lambda)^2 - 2 \tfrac{1}{4} \lambda g - 7 g^2 \right)\bar{\sigma}^2 \bar{\varsigma}^2\nonumber\\
  &\phantom{mmm} 
  - \frac{1}{24} Z_k (\bar{\sigma}^2 + \bar{\varsigma}^2) \Big( 15 u_k (\bar{\sigma}^2 - \bar{\varsigma}^2)^2 (g - \tfrac{1}{4} \lambda)
  + 2 w_k \left( \tfrac{1}{4}\lambda_k (\bar{\sigma}^2 - \bar{\varsigma}^2)^2 - g (\bar{\sigma}^4 + 30\bar{\sigma}^2\bar{\varsigma}^2 + \bar{\varsigma}^4) \right) \Big) \nonumber\\
     &\phantom{mmm} 
  + \frac{1}{2304}Z_k^2 (\bar{\sigma}^2 + \bar{\varsigma}^2)^2 \Big( 
      225 u_k^2 (\bar{\sigma}^2 - \bar{\varsigma}^2)^2 - 60 u_k w_k (\bar{\sigma}^2 - \bar{\varsigma}^2)^2
  + 4 w_k^2 ( \bar{\sigma}^4 + 62\bar{\sigma}^2\bar{\varsigma}^2 + \bar{\varsigma}^4)
  \Big)
    \Big]^{1/2}.
\end{align}
The other eigenvalues are
\begin{align}
\gamma^{A^1(1)}_k &= \frac{Z_{A,k}}{\xi_k} q^2+  Z_k e^2 {\sigma_0}^2, \nonumber\\
\gamma^{A^1(2)}_k &= Z_{A,k} q^2 + Z_k e^2 {\sigma_0}^2   \quad \text{[multiplicity: $d-1$]},\nonumber\\
\gamma^{A^2(1)}_k &= \frac{Z_{A,k}}{\xi_k} q^2+  Z_k e^2 {\varsigma_0}^2,\nonumber\\
\gamma^{A^2(2)}_k &= Z_{A,k} q^2 + Z_k e^2 {\varsigma_0}^2   \quad \text{[multiplicity: $d-1$]}, \nonumber\\
 \gamma^{\sigma_a}_k &= 
 Z_k \left(q^2 + m^2 + \frac{1}{6} Z_k \lambda {\sigma_0}^2 + 2Z_k g {\varsigma_0}^2 +\frac{1}{8} Z_k^2 u_k {\sigma_0}^4 + \frac{1}{12}Z_k^2 w_k {\varsigma_0}^2(2 {\sigma_0}^2 + {\varsigma_0}^2) \right), \quad \text{[multiplicity: $M-1$]},
  \nonumber\\
   \gamma^{\varsigma_a}_k &= 
      Z_k \left(q^2 + m^2 + \frac{1}{6} Z_k \lambda {\varsigma_0}^2+ 2Z_k g {\sigma_0}^2 +\frac{1}{8} Z_k^2 u_k {\varsigma_0}^4 + \frac{1}{12}Z_k^2 w_k {\sigma_0}^2(2 {\varsigma_0}^2 + {\sigma_0}^2)\right), \quad \text{[multiplicity: $M-1$]} ,
 \nonumber\\
  \gamma^{\pi_1}_k &= 
   Z_k \left(q^2 + m^2 +  \frac{1}{6} Z_k \lambda {\sigma_0}^2  + \frac{Z_k}{Z_A} \xi e^2 {\sigma_0}^2  +2Z_k g {\varsigma_0}^2 + \frac{1}{8} Z_k^2 u_k {\sigma_0}^4 + \frac{1}{12} Z_k^2 w_k {\varsigma_0}^2 (2 {\sigma_0}^2 + {\varsigma_0}^2)  \right),\nonumber\\
   \gamma^{\pi_a}_k &=  
  Z_k \left(q^2 + m^2 + \frac{1}{6} Z_k \lambda {\sigma_0}^2 + 2Z_k g {\varsigma_0}^2+ \frac{1}{8} Z_k^2 u_k {\sigma_0}^4 + \frac{1}{12} Z_k^2 w_k {\varsigma_0}^2 (2 {\sigma_0}^2 + {\varsigma_0}^2) \right),
  \quad \text{[multiplicity: $M-1$]} ,
 \nonumber\\
   \gamma^{\varpi_1 }_k &=   
 Z_k \left(q^2 + m^2 +  \frac{1}{6} Z_k \lambda {\varsigma_0}^2  + \frac{Z_k}{Z_A} \xi e^2 {\varsigma_0}^2  +2Z_k g {\sigma_0}^2 + \frac{1}{8} Z_k^2 u_k {\varsigma_0}^4 + \frac{1}{12} Z_k^2 w_k {\sigma_0}^2 (2 {\varsigma_0}^2 + {\sigma_0}^2)  \right),\nonumber\\
  \gamma^{\varpi_a }_k &= 
  Z_k \left(q^2 + m^2 + \frac{1}{6} Z_k \lambda {\varsigma_0}^2 + 2Z_k g {\sigma_0}^2+ \frac{1}{8} Z_k^2 u_k {\varsigma_0}^4 + \frac{1}{12} Z_k^2 w_k {\sigma_0}^2 (2 {\varsigma_0}^2 + {\sigma_0}^2) \right), \quad \text{[multiplicity: $M-1$]},  \nonumber\\
  \gamma^{c^{1*}c^1}_k &= q^2 + \xi e \frac{Z_k}{Z_{A,k}} {\sigma_0}^2,
  \quad \text{[multiplicity: $2$]}, \nonumber\\
  \gamma^{c^{2*}c^2}_k &= q^2 + \xi e \frac{Z_k}{Z_{A,k}} {\varsigma_0}^2 \qquad \text{[multiplicity: $2$]} .
\end{align}
\end{widetext}

\bibliography{RGreferences}

\begin{thebibliography}{36}%
\makeatletter
\providecommand \@ifxundefined [1]{%
 \@ifx{#1\undefined}
}%
\providecommand \@ifnum [1]{%
 \ifnum #1\expandafter \@firstoftwo
 \else \expandafter \@secondoftwo
 \fi
}%
\providecommand \@ifx [1]{%
 \ifx #1\expandafter \@firstoftwo
 \else \expandafter \@secondoftwo
 \fi
}%
\providecommand \natexlab [1]{#1}%
\providecommand \enquote  [1]{``#1''}%
\providecommand \bibnamefont  [1]{#1}%
\providecommand \bibfnamefont [1]{#1}%
\providecommand \citenamefont [1]{#1}%
\providecommand \href@noop [0]{\@secondoftwo}%
\providecommand \href [0]{\begingroup \@sanitize@url \@href}%
\providecommand \@href[1]{\@@startlink{#1}\@@href}%
\providecommand \@@href[1]{\endgroup#1\@@endlink}%
\providecommand \@sanitize@url [0]{\catcode `\\12\catcode `\$12\catcode
  `\&12\catcode `\#12\catcode `\^12\catcode `\_12\catcode `\%12\relax}%
\providecommand \@@startlink[1]{}%
\providecommand \@@endlink[0]{}%
\providecommand \url  [0]{\begingroup\@sanitize@url \@url }%
\providecommand \@url [1]{\endgroup\@href {#1}{\urlprefix }}%
\providecommand \urlprefix  [0]{URL }%
\providecommand \Eprint [0]{\href }%
\providecommand \doibase [0]{https://doi.org/}%
\providecommand \selectlanguage [0]{\@gobble}%
\providecommand \bibinfo  [0]{\@secondoftwo}%
\providecommand \bibfield  [0]{\@secondoftwo}%
\providecommand \translation [1]{[#1]}%
\providecommand \BibitemOpen [0]{}%
\providecommand \bibitemStop [0]{}%
\providecommand \bibitemNoStop [0]{.\EOS\space}%
\providecommand \EOS [0]{\spacefactor3000\relax}%
\providecommand \BibitemShut  [1]{\csname bibitem#1\endcsname}%
\let\auto@bib@innerbib\@empty
\bibitem [{\citenamefont {Herbut}(2007)}]{Herbut07}%
  \BibitemOpen
  \bibfield  {author} {\bibinfo {author} {\bibfnamefont {I.}~\bibnamefont
  {Herbut}},\ }\href {https://doi.org/10.1017/CBO9780511755521} {\emph
  {\bibinfo {title} {A Modern Approach to Critical Phenomena}}}\ (\bibinfo
  {publisher} {Cambridge University Press, Cambridge, England, 2007},\ \bibinfo
  {year} {2007})\BibitemShut {NoStop}%
\bibitem [{\citenamefont {Kleinert}(1983)}]{Kleinert83}%
  \BibitemOpen
  \bibfield  {author} {\bibinfo {author} {\bibfnamefont {H.}~\bibnamefont
  {Kleinert}},\ }\bibfield  {title} {\bibinfo {title} {Limitations to the
  {C}oleman-{W}einberg mechanism of spontaneous mass generation},\ }\href
  {https://doi.org/10.1016/0370-2693(83)90075-8} {\bibfield  {journal}
  {\bibinfo  {journal} {Phys. Lett.}\ }\textbf {\bibinfo {volume} {128B}},\
  \bibinfo {pages} {69} (\bibinfo {year} {1983})}\BibitemShut {NoStop}%
\bibitem [{\citenamefont {Herbut}\ and\ \citenamefont
  {Tesanovic}(1996)}]{Herbut96}%
  \BibitemOpen
  \bibfield  {author} {\bibinfo {author} {\bibfnamefont {I.~F.}\ \bibnamefont
  {Herbut}}\ and\ \bibinfo {author} {\bibfnamefont {Z.}~\bibnamefont
  {Tesanovic}},\ }\bibfield  {title} {\bibinfo {title} {Critical fluctuations
  in superconductors and the magnetic field penetration depth},\ }\href
  {https://doi.org/10.1103/PhysRevLett.76.4588} {\bibfield  {journal} {\bibinfo
   {journal} {Phys. Rev. Lett.}\ }\textbf {\bibinfo {volume} {76}},\ \bibinfo
  {pages} {4588} (\bibinfo {year} {1996})}\BibitemShut {NoStop}%
\bibitem [{\citenamefont {Bergerhoff}\ \emph {et~al.}(1996)\citenamefont
  {Bergerhoff}, \citenamefont {Freire}, \citenamefont {Litim}, \citenamefont
  {Lola},\ and\ \citenamefont {Wetterich}}]{Bergerhoff96}%
  \BibitemOpen
  \bibfield  {author} {\bibinfo {author} {\bibfnamefont {B.}~\bibnamefont
  {Bergerhoff}}, \bibinfo {author} {\bibfnamefont {F.}~\bibnamefont {Freire}},
  \bibinfo {author} {\bibfnamefont {D.~F.}\ \bibnamefont {Litim}}, \bibinfo
  {author} {\bibfnamefont {S.}~\bibnamefont {Lola}},\ and\ \bibinfo {author}
  {\bibfnamefont {C.}~\bibnamefont {Wetterich}},\ }\bibfield  {title} {\bibinfo
  {title} {Phase diagram of superconductors from nonperturbative flow
  equations},\ }\href {https://doi.org/10.1103/PhysRevB.53.5734} {\bibfield
  {journal} {\bibinfo  {journal} {Phys. Rev. B}\ }\textbf {\bibinfo {volume}
  {53}},\ \bibinfo {pages} {5734} (\bibinfo {year} {1996})}\BibitemShut
  {NoStop}%
\bibitem [{\citenamefont {Fejos}\ and\ \citenamefont
  {Hatsuda}(2016)}]{FejosHatsuda16}%
  \BibitemOpen
  \bibfield  {author} {\bibinfo {author} {\bibfnamefont {G.}~\bibnamefont
  {Fejos}}\ and\ \bibinfo {author} {\bibfnamefont {T.}~\bibnamefont
  {Hatsuda}},\ }\bibfield  {title} {\bibinfo {title} {Fixed point structure of
  the {A}belian {H}iggs model},\ }\href
  {https://doi.org/10.1103/PhysRevD.93.121701} {\bibfield  {journal} {\bibinfo
  {journal} {Phys. Rev. D}\ }\textbf {\bibinfo {volume} {93}},\ \bibinfo
  {pages} {121701(R)} (\bibinfo {year} {2016})}\BibitemShut {NoStop}%
\bibitem [{\citenamefont {Fejos}\ and\ \citenamefont
  {Hatsuda}(2017)}]{FejosHatsuda17}%
  \BibitemOpen
  \bibfield  {author} {\bibinfo {author} {\bibfnamefont {G.}~\bibnamefont
  {Fejos}}\ and\ \bibinfo {author} {\bibfnamefont {T.}~\bibnamefont
  {Hatsuda}},\ }\bibfield  {title} {\bibinfo {title} {Renormalization group
  flows of the {$N$}-component {A}belian {H}iggs model},\ }\href
  {https://doi.org/10.1103/PhysRevD.96.056018} {\bibfield  {journal} {\bibinfo
  {journal} {Phys. Rev. D}\ }\textbf {\bibinfo {volume} {96}},\ \bibinfo
  {pages} {056018} (\bibinfo {year} {2017})}\BibitemShut {NoStop}%
\bibitem [{\citenamefont {Kosterlitz}\ \emph {et~al.}(1976)\citenamefont
  {Kosterlitz}, \citenamefont {Nelson},\ and\ \citenamefont
  {Fisher}}]{Kosterlitz76}%
  \BibitemOpen
  \bibfield  {author} {\bibinfo {author} {\bibfnamefont {J.~M.}\ \bibnamefont
  {Kosterlitz}}, \bibinfo {author} {\bibfnamefont {D.~R.}\ \bibnamefont
  {Nelson}},\ and\ \bibinfo {author} {\bibfnamefont {M.~E.}\ \bibnamefont
  {Fisher}},\ }\bibfield  {title} {\bibinfo {title} {Bicritical and
  tetracritical points in anisotropic antiferromagnetic systems},\ }\href
  {https://doi.org/10.1103/PhysRevB.13.412} {\bibfield  {journal} {\bibinfo
  {journal} {Phys. Rev. B}\ }\textbf {\bibinfo {volume} {13}},\ \bibinfo
  {pages} {412} (\bibinfo {year} {1976})}\BibitemShut {NoStop}%
\bibitem [{\citenamefont {Ceccarelli}\ \emph {et~al.}(2015)\citenamefont
  {Ceccarelli}, \citenamefont {Nespolo}, \citenamefont {Pelissetto},\ and\
  \citenamefont {Vicari}}]{Ceccarelli15}%
  \BibitemOpen
  \bibfield  {author} {\bibinfo {author} {\bibfnamefont {G.}~\bibnamefont
  {Ceccarelli}}, \bibinfo {author} {\bibfnamefont {J.}~\bibnamefont {Nespolo}},
  \bibinfo {author} {\bibfnamefont {A.}~\bibnamefont {Pelissetto}},\ and\
  \bibinfo {author} {\bibfnamefont {E.}~\bibnamefont {Vicari}},\ }\bibfield
  {title} {\bibinfo {title} {Bose-{E}instein condensation and critical behavior
  of two-component bosonic gases},\ }\href
  {https://doi.org/10.1103/PhysRevA.92.043613} {\bibfield  {journal} {\bibinfo
  {journal} {Phys. Rev. A}\ }\textbf {\bibinfo {volume} {92}},\ \bibinfo
  {pages} {043613} (\bibinfo {year} {2015})}\BibitemShut {NoStop}%
\bibitem [{\citenamefont {Ceccarelli}\ \emph {et~al.}(2016)\citenamefont
  {Ceccarelli}, \citenamefont {Nespolo}, \citenamefont {Pelissetto},\ and\
  \citenamefont {Vicari}}]{Ceccarelli16}%
  \BibitemOpen
  \bibfield  {author} {\bibinfo {author} {\bibfnamefont {G.}~\bibnamefont
  {Ceccarelli}}, \bibinfo {author} {\bibfnamefont {J.}~\bibnamefont {Nespolo}},
  \bibinfo {author} {\bibfnamefont {A.}~\bibnamefont {Pelissetto}},\ and\
  \bibinfo {author} {\bibfnamefont {E.}~\bibnamefont {Vicari}},\ }\bibfield
  {title} {\bibinfo {title} {Phase diagram and multicritical behaviors of
  mixtures of three-dimensional bosonic gases},\ }\href
  {https://doi.org/10.1103/PhysRevA.93.033647} {\bibfield  {journal} {\bibinfo
  {journal} {Phys. Rev. A}\ }\textbf {\bibinfo {volume} {93}},\ \bibinfo
  {pages} {033647} (\bibinfo {year} {2016})}\BibitemShut {NoStop}%
\bibitem [{\citenamefont {Zaanen}\ \emph {et~al.}(2004)\citenamefont {Zaanen},
  \citenamefont {Nussinov},\ and\ \citenamefont {Mukhin}}]{Zaanen04}%
  \BibitemOpen
  \bibfield  {author} {\bibinfo {author} {\bibfnamefont {J.}~\bibnamefont
  {Zaanen}}, \bibinfo {author} {\bibfnamefont {Z.}~\bibnamefont {Nussinov}},\
  and\ \bibinfo {author} {\bibfnamefont {S.}~\bibnamefont {Mukhin}},\
  }\bibfield  {title} {\bibinfo {title} {Duality in 2+1{D} quantum elasticity:
  Superconductivity and quantum nematic order},\ }\href
  {https://doi.org/10.1016/j.aop.2003.10.003} {\bibfield  {journal} {\bibinfo
  {journal} {Ann. Phys. (Amsterdam)}\ }\textbf {\bibinfo {volume} {310}},\
  \bibinfo {pages} {181} (\bibinfo {year} {2004})}\BibitemShut {NoStop}%
\bibitem [{\citenamefont {Beekman}\ \emph {et~al.}(2017)\citenamefont
  {Beekman}, \citenamefont {Nissinen}, \citenamefont {Wu}, \citenamefont {Liu},
  \citenamefont {Slager}, \citenamefont {Nussinov}, \citenamefont {Cvetkovic},\
  and\ \citenamefont {Zaanen}}]{Beekman17}%
  \BibitemOpen
  \bibfield  {author} {\bibinfo {author} {\bibfnamefont {A.}~\bibnamefont
  {Beekman}}, \bibinfo {author} {\bibfnamefont {J.}~\bibnamefont {Nissinen}},
  \bibinfo {author} {\bibfnamefont {K.}~\bibnamefont {Wu}}, \bibinfo {author}
  {\bibfnamefont {K.}~\bibnamefont {Liu}}, \bibinfo {author} {\bibfnamefont
  {R.-J.}\ \bibnamefont {Slager}}, \bibinfo {author} {\bibfnamefont
  {Z.}~\bibnamefont {Nussinov}}, \bibinfo {author} {\bibfnamefont
  {V.}~\bibnamefont {Cvetkovic}},\ and\ \bibinfo {author} {\bibfnamefont
  {J.}~\bibnamefont {Zaanen}},\ }\bibfield  {title} {\bibinfo {title} {Dual
  gauge field theory of quantum liquid crystals in two dimensions},\ }\href
  {https://doi.org/10.1016/j.physrep.2017.03.004} {\bibfield  {journal}
  {\bibinfo  {journal} {Phys. Rep.}\ }\textbf {\bibinfo {volume} {683}},\
  \bibinfo {pages} {1} (\bibinfo {year} {2017})}\BibitemShut {NoStop}%
\bibitem [{\citenamefont {Nakamura}\ \emph {et~al.}(2016)\citenamefont
  {Nakamura}, \citenamefont {Matsui}, \citenamefont {Matsui},\ and\
  \citenamefont {Fukuyama}}]{Nakamura16}%
  \BibitemOpen
  \bibfield  {author} {\bibinfo {author} {\bibfnamefont {S.}~\bibnamefont
  {Nakamura}}, \bibinfo {author} {\bibfnamefont {K.}~\bibnamefont {Matsui}},
  \bibinfo {author} {\bibfnamefont {T.}~\bibnamefont {Matsui}},\ and\ \bibinfo
  {author} {\bibfnamefont {H.}~\bibnamefont {Fukuyama}},\ }\bibfield  {title}
  {\bibinfo {title} {Possible quantum liquid crystal phases of helium
  monolayers},\ }\href {https://doi.org/10.1103/PhysRevB.94.180501} {\bibfield
  {journal} {\bibinfo  {journal} {Phys. Rev. B}\ }\textbf {\bibinfo {volume}
  {94}},\ \bibinfo {pages} {180501(R)} (\bibinfo {year} {2016})}\BibitemShut
  {NoStop}%
\bibitem [{\citenamefont {Berges}\ \emph {et~al.}(2002)\citenamefont {Berges},
  \citenamefont {Tetradis},\ and\ \citenamefont {Wetterich}}]{Berges02}%
  \BibitemOpen
  \bibfield  {author} {\bibinfo {author} {\bibfnamefont {J.}~\bibnamefont
  {Berges}}, \bibinfo {author} {\bibfnamefont {N.}~\bibnamefont {Tetradis}},\
  and\ \bibinfo {author} {\bibfnamefont {C.}~\bibnamefont {Wetterich}},\
  }\bibfield  {title} {\bibinfo {title} {Non-perturbative renormalization flow
  in quantum field theory and statistical physics},\ }\href
  {https://doi.org/10.1016/S0370-1573(01)00098-9} {\bibfield  {journal}
  {\bibinfo  {journal} {Phys. Rep.}\ }\textbf {\bibinfo {volume} {363}},\
  \bibinfo {pages} {223} (\bibinfo {year} {2002})}\BibitemShut {NoStop}%
\bibitem [{\citenamefont {Kopietz}\ \emph {et~al.}(2010)\citenamefont
  {Kopietz}, \citenamefont {Bartosch},\ and\ \citenamefont
  {Schütz}}]{Kopietz10}%
  \BibitemOpen
  \bibfield  {author} {\bibinfo {author} {\bibfnamefont {P.}~\bibnamefont
  {Kopietz}}, \bibinfo {author} {\bibfnamefont {L.}~\bibnamefont {Bartosch}},\
  and\ \bibinfo {author} {\bibfnamefont {F.}~\bibnamefont {Schütz}},\ }\href
  {https://doi.org/10.1007/978-3-642-05094-7} {\emph {\bibinfo {title}
  {Introduction to the Functional Renormalization Group}}},\ \bibinfo {series}
  {Lecture Notes in Physics}, Vol.\ \bibinfo {volume} {798}\ (\bibinfo
  {publisher} {Springer, New York},\ \bibinfo {year} {2010})\BibitemShut
  {NoStop}%
\bibitem [{\citenamefont {Papenbrock}\ and\ \citenamefont
  {Wetterich}(1995)}]{PapenbrockWetterich95}%
  \BibitemOpen
  \bibfield  {author} {\bibinfo {author} {\bibfnamefont {T.}~\bibnamefont
  {Papenbrock}}\ and\ \bibinfo {author} {\bibfnamefont {C.}~\bibnamefont
  {Wetterich}},\ }\bibfield  {title} {\bibinfo {title} {Two-loop results from
  improved one loop computations},\ }\href {https://doi.org/10.1007/BF01556140}
  {\bibfield  {journal} {\bibinfo  {journal} {Z. Phys. C}\ }\textbf {\bibinfo
  {volume} {65}},\ \bibinfo {pages} {519} (\bibinfo {year} {1995})}\BibitemShut
  {NoStop}%
\bibitem [{\citenamefont {Eichhorn}\ \emph {et~al.}(2013)\citenamefont
  {Eichhorn}, \citenamefont {Mesterh\'azy},\ and\ \citenamefont
  {Scherer}}]{Eichhorn13}%
  \BibitemOpen
  \bibfield  {author} {\bibinfo {author} {\bibfnamefont {A.}~\bibnamefont
  {Eichhorn}}, \bibinfo {author} {\bibfnamefont {D.}~\bibnamefont
  {Mesterh\'azy}},\ and\ \bibinfo {author} {\bibfnamefont {M.~M.}\ \bibnamefont
  {Scherer}},\ }\bibfield  {title} {\bibinfo {title} {Multicritical behavior in
  models with two competing order parameters},\ }\href
  {https://doi.org/10.1103/PhysRevE.88.042141} {\bibfield  {journal} {\bibinfo
  {journal} {Phys. Rev. E}\ }\textbf {\bibinfo {volume} {88}},\ \bibinfo
  {pages} {042141} (\bibinfo {year} {2013})}\BibitemShut {NoStop}%
\bibitem [{\citenamefont {Abreu}\ \emph {et~al.}(2008)\citenamefont {Abreu},
  \citenamefont {de~Calan},\ and\ \citenamefont {Malbouisson}}]{Abreu08}%
  \BibitemOpen
  \bibfield  {author} {\bibinfo {author} {\bibfnamefont {L.}~\bibnamefont
  {Abreu}}, \bibinfo {author} {\bibfnamefont {C.}~\bibnamefont {de~Calan}},\
  and\ \bibinfo {author} {\bibfnamefont {A.}~\bibnamefont {Malbouisson}},\
  }\bibfield  {title} {\bibinfo {title} {Multicritical behavior of the
  two-field {G}inzburg-{L}andau model coupled to a gauge field},\ }\href
  {https://doi.org/10.1016/j.physa.2007.09.033} {\bibfield  {journal} {\bibinfo
   {journal} {Physica (Amsterdam)}\ }\textbf {\bibinfo {volume} {387A}},\
  \bibinfo {pages} {817} (\bibinfo {year} {2008})}\BibitemShut {NoStop}%
\bibitem [{\citenamefont {Sakhi}(2013)}]{Sakhi13}%
  \BibitemOpen
  \bibfield  {author} {\bibinfo {author} {\bibfnamefont {S.}~\bibnamefont
  {Sakhi}},\ }\bibfield  {title} {\bibinfo {title} {Quantum critical properties
  in the topological {G}inzburg-{L}andau theory of self-dual {J}osephson
  junction arrays},\ }\href {https://doi.org/10.1016/j.physa.2013.08.009}
  {\bibfield  {journal} {\bibinfo  {journal} {Physica (Amsterdam)}\ }\textbf
  {\bibinfo {volume} {392A}},\ \bibinfo {pages} {6255} (\bibinfo {year}
  {2013})}\BibitemShut {NoStop}%
\bibitem [{\citenamefont {Sakhi}(2018)}]{Sakhi18}%
  \BibitemOpen
  \bibfield  {author} {\bibinfo {author} {\bibfnamefont {S.}~\bibnamefont
  {Sakhi}},\ }\bibfield  {title} {\bibinfo {title} {Tricritical behavior in the
  {C}hern-{S}imons-{G}inzburg-{L}andau theory of self-dual {J}osephson junction
  arrays},\ }\href {https://doi.org/10.1103/PhysRevD.97.096015} {\bibfield
  {journal} {\bibinfo  {journal} {Phys. Rev. D}\ }\textbf {\bibinfo {volume}
  {97}},\ \bibinfo {pages} {096015} (\bibinfo {year} {2018})}\BibitemShut
  {NoStop}%
\bibitem [{\citenamefont {Pelissetto}\ and\ \citenamefont
  {Vicari}(2002)}]{Pelissetto02}%
  \BibitemOpen
  \bibfield  {author} {\bibinfo {author} {\bibfnamefont {A.}~\bibnamefont
  {Pelissetto}}\ and\ \bibinfo {author} {\bibfnamefont {E.}~\bibnamefont
  {Vicari}},\ }\bibfield  {title} {\bibinfo {title} {Critical phenomena and
  renormalization-group theory},\ }\href
  {https://doi.org/10.1016/S0370-1573(02)00219-3} {\bibfield  {journal}
  {\bibinfo  {journal} {Phys. Rep.}\ }\textbf {\bibinfo {volume} {368}},\
  \bibinfo {pages} {549} (\bibinfo {year} {2002})}\BibitemShut {NoStop}%
\bibitem [{\citenamefont {Pelissetto}\ and\ \citenamefont
  {Vicari}(2005)}]{Pelissetto05}%
  \BibitemOpen
  \bibfield  {author} {\bibinfo {author} {\bibfnamefont {A.}~\bibnamefont
  {Pelissetto}}\ and\ \bibinfo {author} {\bibfnamefont {E.}~\bibnamefont
  {Vicari}},\ }\bibfield  {title} {\bibinfo {title} {Interacting {$N$}-vector
  order parameters with {$O(N)$} symmetry},\ }\href
  {https://doi.org/10.5488/CMP.8.1.87} {\bibfield  {journal} {\bibinfo
  {journal} {Condens. Matter Phys.}\ }\textbf {\bibinfo {volume} {8}},\
  \bibinfo {pages} {87} (\bibinfo {year} {2005})}\BibitemShut {NoStop}%
\bibitem [{\citenamefont {Pismak}\ \emph {et~al.}(2009)\citenamefont {Pismak},
  \citenamefont {Weber},\ and\ \citenamefont {Wegner}}]{Pismak09}%
  \BibitemOpen
  \bibfield  {author} {\bibinfo {author} {\bibfnamefont {Y.~M.}\ \bibnamefont
  {Pismak}}, \bibinfo {author} {\bibfnamefont {A.}~\bibnamefont {Weber}},\ and\
  \bibinfo {author} {\bibfnamefont {F.~J.}\ \bibnamefont {Wegner}},\ }\bibfield
   {title} {\bibinfo {title} {Critical behavior of a general
  $\mathrm{O}{(n)}$-symmetric model of two $n$-vector fields in {$D= 4 -
  2\epsilon$}},\ }\href {https://doi.org/10.1088/1751-8113/42/9/095003}
  {\bibfield  {journal} {\bibinfo  {journal} {J. Phys. A}\ }\textbf {\bibinfo
  {volume} {42}},\ \bibinfo {pages} {095003} (\bibinfo {year}
  {2009})}\BibitemShut {NoStop}%
\bibitem [{\citenamefont {Calabrese}\ \emph {et~al.}(2003)\citenamefont
  {Calabrese}, \citenamefont {Pelissetto},\ and\ \citenamefont
  {Vicari}}]{Calabrese03}%
  \BibitemOpen
  \bibfield  {author} {\bibinfo {author} {\bibfnamefont {P.}~\bibnamefont
  {Calabrese}}, \bibinfo {author} {\bibfnamefont {A.}~\bibnamefont
  {Pelissetto}},\ and\ \bibinfo {author} {\bibfnamefont {E.}~\bibnamefont
  {Vicari}},\ }\bibfield  {title} {\bibinfo {title} {Multicritical phenomena in
  $\mathrm{O}{(n}_{1})\ensuremath{\bigoplus}\mathrm{O}{(n}_{2})$-symmetric
  theories},\ }\href {https://doi.org/10.1103/PhysRevB.67.054505} {\bibfield
  {journal} {\bibinfo  {journal} {Phys. Rev. B}\ }\textbf {\bibinfo {volume}
  {67}},\ \bibinfo {pages} {054505} (\bibinfo {year} {2003})}\BibitemShut
  {NoStop}%
\bibitem [{\citenamefont {Calabrese}\ \emph {et~al.}(2004)\citenamefont
  {Calabrese}, \citenamefont {Parruccini}, \citenamefont {Pelissetto},\ and\
  \citenamefont {Vicari}}]{Calabrese04}%
  \BibitemOpen
  \bibfield  {author} {\bibinfo {author} {\bibfnamefont {P.}~\bibnamefont
  {Calabrese}}, \bibinfo {author} {\bibfnamefont {P.}~\bibnamefont
  {Parruccini}}, \bibinfo {author} {\bibfnamefont {A.}~\bibnamefont
  {Pelissetto}},\ and\ \bibinfo {author} {\bibfnamefont {E.}~\bibnamefont
  {Vicari}},\ }\bibfield  {title} {\bibinfo {title} {Critical behavior of
  $\mathrm{O}(2)\ensuremath{\bigotimes}\mathrm{O}(n)$ symmetric models},\
  }\href {https://doi.org/10.1103/PhysRevB.70.174439} {\bibfield  {journal}
  {\bibinfo  {journal} {Phys. Rev. B}\ }\textbf {\bibinfo {volume} {70}},\
  \bibinfo {pages} {174439} (\bibinfo {year} {2004})}\BibitemShut {NoStop}%
\bibitem [{\citenamefont {Bornholdt}\ \emph {et~al.}(1995)\citenamefont
  {Bornholdt}, \citenamefont {Tetradis},\ and\ \citenamefont
  {Wetterich}}]{Bornholdt95}%
  \BibitemOpen
  \bibfield  {author} {\bibinfo {author} {\bibfnamefont {S.}~\bibnamefont
  {Bornholdt}}, \bibinfo {author} {\bibfnamefont {N.}~\bibnamefont
  {Tetradis}},\ and\ \bibinfo {author} {\bibfnamefont {C.}~\bibnamefont
  {Wetterich}},\ }\bibfield  {title} {\bibinfo {title} {{C}oleman-{W}einberg
  phase transition in two-scalar models},\ }\href
  {https://doi.org/10.1016/0370-2693(95)00045-M} {\bibfield  {journal}
  {\bibinfo  {journal} {Phys. Lett. B}\ }\textbf {\bibinfo {volume} {348}},\
  \bibinfo {pages} {89} (\bibinfo {year} {1995})}\BibitemShut {NoStop}%
\bibitem [{\citenamefont {Bornholdt}\ \emph {et~al.}(1996)\citenamefont
  {Bornholdt}, \citenamefont {Tetradis},\ and\ \citenamefont
  {Wetterich}}]{Bornholdt96}%
  \BibitemOpen
  \bibfield  {author} {\bibinfo {author} {\bibfnamefont {S.}~\bibnamefont
  {Bornholdt}}, \bibinfo {author} {\bibfnamefont {N.}~\bibnamefont
  {Tetradis}},\ and\ \bibinfo {author} {\bibfnamefont {C.}~\bibnamefont
  {Wetterich}},\ }\bibfield  {title} {\bibinfo {title} {High temperature phase
  transition in two-scalar theories},\ }\href
  {https://doi.org/10.1103/PhysRevD.53.4552} {\bibfield  {journal} {\bibinfo
  {journal} {Phys. Rev. D}\ }\textbf {\bibinfo {volume} {53}},\ \bibinfo
  {pages} {4552} (\bibinfo {year} {1996})}\BibitemShut {NoStop}%
\bibitem [{\citenamefont {Borchardt}\ and\ \citenamefont
  {Eichhorn}(2016)}]{Borchardt16}%
  \BibitemOpen
  \bibfield  {author} {\bibinfo {author} {\bibfnamefont {J.}~\bibnamefont
  {Borchardt}}\ and\ \bibinfo {author} {\bibfnamefont {A.}~\bibnamefont
  {Eichhorn}},\ }\bibfield  {title} {\bibinfo {title} {Universal behavior of
  coupled order parameters below three dimensions},\ }\href
  {https://doi.org/10.1103/PhysRevE.94.042105} {\bibfield  {journal} {\bibinfo
  {journal} {Phys. Rev. E}\ }\textbf {\bibinfo {volume} {94}},\ \bibinfo
  {pages} {042105} (\bibinfo {year} {2016})}\BibitemShut {NoStop}%
\bibitem [{\citenamefont {Benvenuti}\ and\ \citenamefont
  {Khachatryan}(2018)}]{Benvenuti18}%
  \BibitemOpen
  \bibfield  {author} {\bibinfo {author} {\bibfnamefont {S.}~\bibnamefont
  {Benvenuti}}\ and\ \bibinfo {author} {\bibfnamefont {H.}~\bibnamefont
  {Khachatryan}},\ }\bibfield  {title} {\bibinfo {title} {{QED}'s in $2{+}1$
  dimensions: Complex fixed points and dualities},\ }\Eprint
  {https://arxiv.org/abs/1812.01544} {arXiv:1812.01544 [hep-th]}  (\bibinfo
  {year} {2018})\BibitemShut {NoStop}%
\bibitem [{\citenamefont {Benvenuti}\ and\ \citenamefont
  {Khachatryan}(2019)}]{Benvenuti19}%
  \BibitemOpen
  \bibfield  {author} {\bibinfo {author} {\bibfnamefont {S.}~\bibnamefont
  {Benvenuti}}\ and\ \bibinfo {author} {\bibfnamefont {H.}~\bibnamefont
  {Khachatryan}},\ }\bibfield  {title} {\bibinfo {title} {Easy-plane
  {QED}$_3$'s in the large {$N_f$} limit},\ }\Eprint
  {https://arxiv.org/abs/1902.05767} {arXiv:1902.05767 [hep-th]}  (\bibinfo
  {year} {2019})\BibitemShut {NoStop}%
\bibitem [{\citenamefont {Litim}(2000)}]{Litim00}%
  \BibitemOpen
  \bibfield  {author} {\bibinfo {author} {\bibfnamefont {D.~F.}\ \bibnamefont
  {Litim}},\ }\bibfield  {title} {\bibinfo {title} {Optimisation of the exact
  renormalisation group},\ }\href
  {https://doi.org/10.1016/S0370-2693(00)00748-6} {\bibfield  {journal}
  {\bibinfo  {journal} {Phys. Lett. B}\ }\textbf {\bibinfo {volume} {486}},\
  \bibinfo {pages} {92 } (\bibinfo {year} {2000})}\BibitemShut {NoStop}%
\bibitem [{\citenamefont {Wetterich}(1993)}]{Wetterich93}%
  \BibitemOpen
  \bibfield  {author} {\bibinfo {author} {\bibfnamefont {C.}~\bibnamefont
  {Wetterich}},\ }\bibfield  {title} {\bibinfo {title} {Exact evolution
  equation for the effective potential},\ }\href@noop {} {\bibfield  {journal}
  {\bibinfo  {journal} {Phys. Lett. B}\ }\textbf {\bibinfo {volume} {301}},\
  \bibinfo {pages} {90} (\bibinfo {year} {1993})}\BibitemShut {NoStop}%
\bibitem [{\citenamefont {Bartosch}(2013)}]{Bartosch13}%
  \BibitemOpen
  \bibfield  {author} {\bibinfo {author} {\bibfnamefont {L.}~\bibnamefont
  {Bartosch}},\ }\bibfield  {title} {\bibinfo {title} {Corrections to scaling
  in the critical theory of deconfined criticality},\ }\href
  {https://doi.org/10.1103/PhysRevB.88.195140} {\bibfield  {journal} {\bibinfo
  {journal} {Phys. Rev. B}\ }\textbf {\bibinfo {volume} {88}},\ \bibinfo
  {pages} {195140} (\bibinfo {year} {2013})}\BibitemShut {NoStop}%
\bibitem [{\citenamefont {Kiometzis}\ \emph {et~al.}(1995)\citenamefont
  {Kiometzis}, \citenamefont {Kleinert},\ and\ \citenamefont
  {Schakel}}]{Kiometzis95}%
  \BibitemOpen
  \bibfield  {author} {\bibinfo {author} {\bibfnamefont {M.}~\bibnamefont
  {Kiometzis}}, \bibinfo {author} {\bibfnamefont {H.}~\bibnamefont
  {Kleinert}},\ and\ \bibinfo {author} {\bibfnamefont {A.~M.}\ \bibnamefont
  {Schakel}},\ }\bibfield  {title} {\bibinfo {title} {Dual description of the
  superconducting phase transition},\ }\href
  {https://doi.org/10.1002/prop.2190430803} {\bibfield  {journal} {\bibinfo
  {journal} {Fortschr. Phys.}\ }\textbf {\bibinfo {volume} {43}},\ \bibinfo
  {pages} {697} (\bibinfo {year} {1995})}\BibitemShut {NoStop}%
\bibitem [{\citenamefont {Hove}\ \emph {et~al.}(2000)\citenamefont {Hove},
  \citenamefont {Mo},\ and\ \citenamefont {Sudb\o{}}}]{Hove00}%
  \BibitemOpen
  \bibfield  {author} {\bibinfo {author} {\bibfnamefont {J.}~\bibnamefont
  {Hove}}, \bibinfo {author} {\bibfnamefont {S.}~\bibnamefont {Mo}},\ and\
  \bibinfo {author} {\bibfnamefont {A.}~\bibnamefont {Sudb\o{}}},\ }\bibfield
  {title} {\bibinfo {title} {{H}ausdorff dimension of critical fluctuations in
  {A}belian gauge theories},\ }\href
  {https://doi.org/10.1103/PhysRevLett.85.2368} {\bibfield  {journal} {\bibinfo
   {journal} {Phys. Rev. Lett.}\ }\textbf {\bibinfo {volume} {85}},\ \bibinfo
  {pages} {2368} (\bibinfo {year} {2000})}\BibitemShut {NoStop}%
\bibitem [{\citenamefont {Kajantie}\ \emph {et~al.}(2004)\citenamefont
  {Kajantie}, \citenamefont {Laine}, \citenamefont {Neuhaus}, \citenamefont
  {Rajantie},\ and\ \citenamefont {Rummukainen}}]{Kajantie04}%
  \BibitemOpen
  \bibfield  {author} {\bibinfo {author} {\bibfnamefont {K.}~\bibnamefont
  {Kajantie}}, \bibinfo {author} {\bibfnamefont {M.}~\bibnamefont {Laine}},
  \bibinfo {author} {\bibfnamefont {T.}~\bibnamefont {Neuhaus}}, \bibinfo
  {author} {\bibfnamefont {A.}~\bibnamefont {Rajantie}},\ and\ \bibinfo
  {author} {\bibfnamefont {K.}~\bibnamefont {Rummukainen}},\ }\bibfield
  {title} {\bibinfo {title} {Duality and scaling in 3-dimensional scalar
  electrodynamics},\ }\href {https://doi.org/10.1016/j.nuclphysb.2004.08.018}
  {\bibfield  {journal} {\bibinfo  {journal} {Nucl. Phys.}\ }\textbf {\bibinfo
  {volume} {B699}},\ \bibinfo {pages} {632 } (\bibinfo {year}
  {2004})}\BibitemShut {NoStop}%
\bibitem [{\citenamefont {Nahum}\ and\ \citenamefont
  {Chalker}(2012)}]{Nahum12}%
  \BibitemOpen
  \bibfield  {author} {\bibinfo {author} {\bibfnamefont {A.}~\bibnamefont
  {Nahum}}\ and\ \bibinfo {author} {\bibfnamefont {J.~T.}\ \bibnamefont
  {Chalker}},\ }\bibfield  {title} {\bibinfo {title} {Universal statistics of
  vortex lines},\ }\href {https://doi.org/10.1103/PhysRevE.85.031141}
  {\bibfield  {journal} {\bibinfo  {journal} {Phys. Rev. E}\ }\textbf {\bibinfo
  {volume} {85}},\ \bibinfo {pages} {031141} (\bibinfo {year}
  {2012})}\BibitemShut {NoStop}%
\end{thebibliography}%
 
\end{document}